\documentclass[fleqn,usenatbib]{mnras}

\usepackage{newtxtext,newtxmath}
\usepackage[T1]{fontenc}
\usepackage{ae,aecompl}

\usepackage{graphicx}	% Including figure files
\usepackage{amsmath} % or simply amstext
\usepackage{pdflscape}
\usepackage{longtable}
\usepackage{caption}
\usepackage{xtab}
\usepackage{soul}
\usepackage{array}
\usepackage{ragged2e}
\newcommand{\etal}[1]{et al. \,}

\newlength{\rlength}

\title[OuterHalo Stars]{Chemo-dynamics of outer halo dwarf stars, including  \textit{Gaia}-Sausage and \textit{Gaia}-Sequoia candidates}

\author[S. Monty et al.]{Stephanie Monty$^{1,2}$\thanks{E-mail: Stephanie.Monty@anu.edu.au},
Kim. A. Venn$^{1}$, 
James M. M. Lane$^{1,3}$, 
Deborah Lokhorst$^{1,3}$,
\newauthor
David Yong$^{2}$
\\
$^{1}$Department of Physics and Astronomy, University of Victoria, Victoria, BC V8W 3P2, Canada\\
$^{2}$Research School of Astronomy and Astrophysics, Australian National University, Canberra, ACT 2611, Australia\\
$^{3}$Department of Astronomy and Astrophysics, University of Toronto, 50 St. George Street, Toronto, ON M5S 3H4, Canada\\
}

\date{Accepted XXX. Received YYY; in original form ZZZ}

\pubyear{2020}
\begin{document}
\label{firstpage}
% \pagerange{\pageref{firstpage}--\pageref{lastpage}}
\maketitle

\begin{abstract}
The low-metallicity, kinematically interesting dwarf stars studied by Stephens \& Boesgaard (2002, SB02)
are re-examined using Gaia DR2 astrometry, and updated model atmospheres and atomic line data. New stellar parameters are determined based on the Gaia DR2 parallactic distances and Dartmouth Stellar Evolution Database isochrones.  
These are in excellent agreement with spectroscopically determined stellar parameters for stars with [Fe/H]$>-2$; however, large disagreements are found for stars with [Fe/H]$\le-2$, with offsets as large as $\Delta$T$_{\rm eff}\sim+500$ K and $\Delta$log\,$g\sim+1.0$.
A subset of six stars (test cases) are analysed ab initio using high resolution spectra with Keck HIRES and Gemini GRACES.
This sub-sample is found to include two $\alpha$-challenged dwarf stars, suggestive of origins in a low mass, accreted dwarf galaxy. The orbital parameters for the entire SB02 sample are re-determined using \textit{Gaia} DR2 data. We find 11 stars that are dynamically coincident with the \textit{Gaia}-Sausage accretion event and another 17 with the \textit{Gaia}-Sequoia event in action space. Both associations include low-mass, metal-poor stars with isochrone ages older than 10 Gyr. Two dynamical subsets are identified within \textit{Gaia}-Sequoia.  When these subsets are examined separately, a common knee in [$\alpha$/Fe] is found for the \textit{Gaia}-Sausage and low orbital energy \textit{Gaia}-Sequoia stars.  A lower metallicity knee is tentatively identified in the \textit{Gaia}-Sequoia high orbital energy stars. If the metal-poor dwarf stars in these samples are true members of the \textit{Gaia}-Sausage and \textit{Gaia}-Sequoia events, then they present a unique opportunity to probe the earlier star formation histories of these systems.
\end{abstract}


%%%%%%%%%%%%%%%%%%%%%%%%%%%%%%%%%%%%%%%%%%%%%%%%%%

%%%%%%%%%%%%%%%%% BODY OF PAPER %%%%%%%%%%%%%%%%%%

\section{Introduction}
Our view of the stars in the solar neighbourhood has drastically changed in the past couple years with the  second data release (DR2) of the European Space Agency's Gaia mission \citep{gaiadr1,gaiadr2}.  The superb astrometric parameters, radial velocities, and photometric data of the billion-star-dataset have impacted fundamental stellar parameter determinations, yielded the first detailed orbits of nearby stars, and revealed a wide variety of structures and streams not previously identified in our Galaxy.

One of the most interesting \textit{Gaia} results thus far has been the discovery of two parallel colour sequences in the Gaia HR diagram of stars in the solar neighbourhood \citep{babusiaux18}. Although they are currently found within 2.5 kpc of the solar neighborhood, these stars have high total or tangential velocities indicating that they belong to the Milky Way (MW) halo. Both \cite{helmige} and \cite{haywoodge} cross matched the dual sequence stars with the SDSS APOGEE database \citep{majewski17} and the \cite{niss10} high orbital energy stars to find that one of the sequences (their ``blue sequence'') is dominated by stars with lower [$\alpha$/Fe] ratios.  
Furthermore, a subset of the stars on the blue sequence are highly retrograde (V $< -500$ kms$^{-1}$) and found in a flattened disk.  Both papers suggest that the blue sequence is dominated by stars from an accreted satellite galaxy with a unique star formation history and chemical evolution (or possibly multiple satellite mergers). 

The idea that accreted stars dominate the low [$\alpha$/Fe] sequence was also proposed by \cite[][NS10/11]{niss10, niss11}, using a smaller sample of stars, and by \cite{belokurov18} using the distribution of many RR Lyrae stars in period-amplitude space. Both \cite{belokurovsausage} and \cite{helmige} have proposed single, but distinctly different merger events $\sim$10 Gyr ago to explain the formation of the MW inner halo, the ``\textit{Gaia}-Sausage'' and ``\textit{Gaia}-Enceladus'' respectively. This is consistent with \cite{gallart19} who showed that the blue sequence stars are identical in age to the red sequence stars through building the star formation histories of both. Alternatively, \cite{myeong19} proposed two merger events, the same ``\textit{Gaia}-Sausage'' event to explain weakly prograde, highly eccentric halo stars and an additional event termed ``\textit{Gaia}-Sequoia'' to explain moderately eccentric, strongly retrograde halo stars. \cite{helmige} and \cite{haywoodge} also suggest that the red sequence is consistent with the Galactic thick disk stars, and that the velocity distribution could be due to dynamical heating of the pre-existing disk by the merger. This had also been proposed \citep[e.g.,][]{gil85,jfn2011} in earlier studies of the chemo-dynamical trends with height from the MW disk. 

Stars in the solar neighbourhood with distinctly different chemical abundances in high energy orbits have been known for nearly two decades \citep[e.g.,][]{fulbright02,sb02,venn04,niss10,hawkins15,bat17}.  The overwhelming majority of stars with halo kinematics have high [$\alpha$/Fe] $\sim$ +0.4, suggesting that they formed in regions with a high star formation rates such that only massive stars and Type II SNe contributed to their chemical enrichment.  Stars with lower [$\alpha$/Fe] $\le$ 0.2 dex tend to be intermediate metallicity stars (with $-1.5 \le$ [Fe/H] $\le -0.5$) that show a slightly declining trend in [$\alpha$/Fe] with increasing [Fe/H]. This chemical pattern is thought to be the result of a slower chemical evolution, with contributions from type Ia supernovae and/or the result of fewer high mass stars in the region
\citep[e.g., a truncated upper initial mass function, IMF,][]{tolstoy03,mcwilliam13}.  

The high-$\alpha$ and low-$\alpha$ sequences are also traced in a number of other elements \citep[Cu, Zn, Y, Ba, Na, Al, Ni;][]{niss11,hawkins15}. The lower abundances of these additional elements has been attributed to larger contributions from metal-poor asymptotic giant branch (AGB) stars, consistent with slower star formation rates and chemical evolution. In their two studies, \cite{fernandez2015, fernandez2017} examined trends in Ca, Mg, and Fe in MW halo stars as a function of galactocentric distance using SDSS DR10 \citep[$R\leq80$ kpc,][]{sdss10} and APOGEE DR12 data \citep[$R\leq30$ kpc,][]{majewski17}. Overall, they found the median [$\alpha$/Fe] abundance was lower by a modest $\sim$0.1 dex for halo stars with [Fe/H] $\sim -1.1$ at distances of R$_{GC}$ $\ge$ 15 kpc, confirming that lower $\alpha$ stars are found at large galactocentric distances.
 
Independent of global trends in the MW halo, a few metal-poor stars with very distinctive chemical abundances have also been found.  In their study, \cite{bat17} derive chemical abundances from high resolution optical spectroscopy of 28 red giant stars in the outer halo ($R_{\text{apo}}$>15 kpc from the galactic centre and height $Z_{\text{max}}>9$ kpc from the MW mid-plane) to examine the halo's chemical diversity.  They find that while the metallicity of the stars in their sample ranges from $-3.1 <$ [Fe/H] $< -0.6$, the [$\alpha$/Fe] abundances remain high across all metallicities, similar to stars in the solar neighbourhood (with the exception of one star anomalously low in [(Ca,Mg)/Fe] $\le -0.2$ that is associated with the Sagittarius stream).  Although they do not find the [$\alpha$/Fe] signature of accretion, they interpret the high values of [Ba/Fe] and [Ba/Y] they find in the intermediate metallicity range stars ([Fe/H]$\sim -1.5$) as evidence for accretion in the outer halo. The high values they find are relative to inner halo stars of the same metallicity and indicate pollution from metal-poor AGB stars. Only one of the stars in their study has measurements in the APOGEE database, but it shows good agreement for elements in common.  

In an earlier study, \cite{ivan03} found three distinct metal-poor dwarf/sub-giant halo stars with [Fe/H] $\sim-2$ and low [$\alpha$/Fe] using high resolution optical spectroscopic data. They showed that each star has additional unique chemical characteristics, such as enhancements in iron-group elements, with one star in their study, CS 22966-043 showing an enormous abundance of [Ga/Fe] = +1.75 (LTE). They suggest that these chemically peculiar stars could have formed in regions enhanced in SN Ia products, similar to the interpretations of the chemical peculiarities of stars in the Carina and Sextans dwarf galaxies \citep{venn12, jablonka2015, nor17}. Clearly our picture of the MW halo is complex, both chemically and kinematically.

Upon examining outer halo stars in the literature, we noticed that a sample of outer halo {\it dwarf} stars studied by \cite[][SB02]{sb02} had not been revisited. The SB02 sample is interesting because the sample includes stars more metal-poor  ($-3.5 <$ [Fe/H] $<-1.5$) than the sample of NS10/11 with large apocentric radii ($R_{\text{apo}}$>15 kpc) and slightly lower [$\alpha$/Fe] $\sim$ 0.1 - 0.2 dex than the majority of halo stars of similar metallicities.
In the original study, SB02 conclude that the stars in their sample do not carry the chemical signatures of an accreted population, thereby forming in situ, in localized MW star forming regions and birthed into orbits reflecting early halo kinematics. Coupling information from the Gaia DR2 data release with significant improvements in stellar spectral analyses, we revisit the chemo-dynamic analysis of the SB02 stars.

The paper is organised with a discussion of the results throughout. In Section \ref{sec:sb02targets} we introduce a subset of the SB02 data re-analysed in this work, briefly describing the techniques used to re-determine stellar parameters, and comparing to the original SB02 parameters. In Section \ref{sec:stellabund} our updated stellar abundances are discussed in the context of both the original SB02 study and the study of NS10/11. In Section \ref{sec:orb} the orbital parameters are updated using Gaia DR2 data, and we discuss the potential origins of the stars in the SB02 sample and accretion history of the Galaxy, highlighting the combination of the chemical and dynamical results. Section \ref{sec:conc} summarises the key results of the paper. An appendix is also included where we discuss techniques used throughout this study and lessons learned as a result.
 
\section{The SB02 Targets}\label{sec:sb02targets} 
\cite{sb02} selected their original sample from the \cite{car94} catalogue of high proper motion stars. The \cite{car94} catalogue was compiled over many years using photometry and radial velocities for almost 500 stars from the Lowell Proper Motion Catalog along with estimates of stellar distances. Orbital parameters including, apocentric radii ($R_{\text{apo}}$) and maximum height from the disc ($Z_{\text{max}}$), were determined in the catalogue using the two component MW model of \citep{bahcall83}.  
SB02 selected 56 dwarf stars that satisfy one of three orbital criterion:\\

$\bullet$ Outer halo ($R_{\text{apo}}$ $> 16$ kpc),

$\bullet$ High halo ($Z_{\text{max}}$ $> 5$ kpc), or

$\bullet$ Extreme retrograde orbit ($V < -400$ km/s).\\
 
In the following section, we re-examine these 56 stars given the new Gaia DR2 astrometry.  In particular, we calculate new stellar parameters by combining the Gaia DR2 parallactic distances with isochrones from the Dartmouth Stellar Evolution Database \citep[\texttt{DSED}, ][]{dsed}.  This is done using an \textit{isochrone-mapping} method, fully described in Appendix \ref{sec:isochronemap}. The new isochrone-derived stellar parameters are then compared to the original spectroscopically determined parameters.

\begin{figure}
\centering
\includegraphics[width=\linewidth]{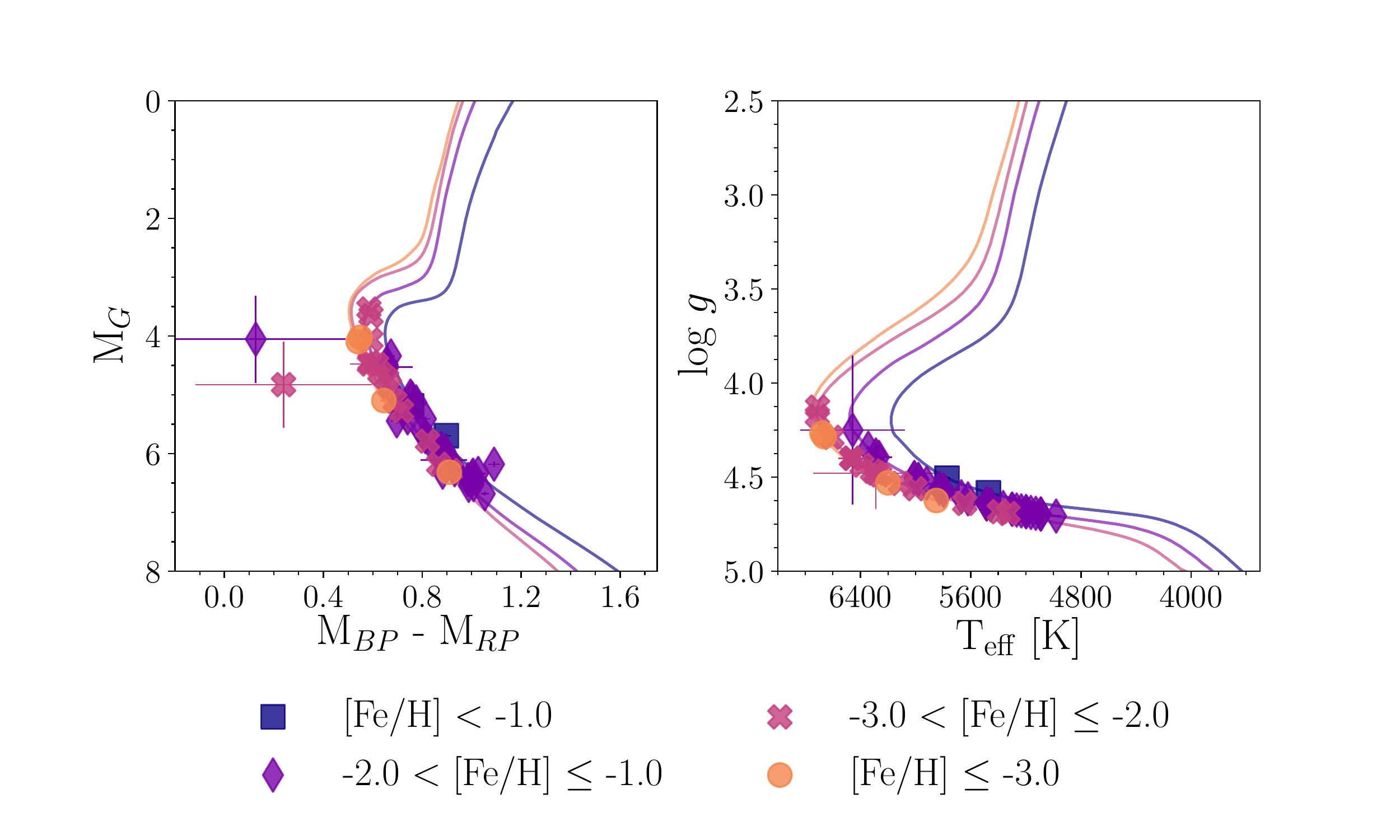}
\caption{\label{fig:sb02_iso} 
Stellar parameters from the \textit{isochrone-mapping} method for the entire Gaia-SB02 sample. See Section \ref{sec:sb02gaia} and Appendix \ref{sec:isochronemap} for details and sample divisions.}
\end{figure} 

\subsection{\label{sec:sb02gaia}SB02 in the Era of Gaia}
Stellar parameters are determined for 54/56 stars in the SB02 sample using our \textit{isochrone-mapping} method (see Appendix \ref{sec:isochronemap}). Two stars in the SB02 sample were removed as they did not have complete photometric information or were unavailable in the Gaia DR2 archive\footnote{\url{https://gea.esac.esa.int/archive/}}. Errors associated with the final stellar parameters are determined using a Monte-Carlo approach to explore the photometric and parallactic parameter uncertainties. Prior to constructing the
Gaia DR2 $G$ vs $BP-RP$ colour-magnitude diagrams (CMDs), we split the SB02 sample into smaller sub-samples using the same metallicity bins used in SB02. Those bins are as follows:\\

$\bullet$ (i) [Fe/H]$>-1.0$, [$\alpha$/Fe] = $0$,

$\bullet$ (ii) $-2.0<$[Fe/H]$\leq-1.0$, [$\alpha$/Fe] = $0.2$,

$\bullet$ (iii) $-3.0<$[Fe/H]$\leq-2.0$, [$\alpha$/Fe] = $0.4$

$\bullet$ (iv) [Fe/H]$\leq-3.0$, [$\alpha$/Fe] = $0.4$. \\

A \texttt{DSED} isochrone \citep{dsed} was created for each of the chemical bins, assuming a fixed age of of 12 Gyr for every bin, the median metallicity for bins (ii) and (iii), and the average metallicity for bins (i) and (iv), respectively. Alpha abundances were adopted for each bin as shown above. The effects of assuming a priori metallicities and alpha abundances for each bin had minimal impact on the stellar parameters, as described in Section \ref{sec:isochronemap}. The results of applying the \textit{isochrone-mapping} method to the entire SB02 sample are shown in Fig.~\ref{fig:sb02_iso}. Thanks in part to the exquisite Gaia DR2 \citep{gaiadr2} data for these stars and their close proximity, it is clear from Fig.~\ref{fig:sb02_iso} that there are no dwarf-giant degenerate solutions for any of the stars. The two outlying points from the second and third bins seen in Fig.~\ref{fig:sb02_iso} have large associated uncertainties in reddening (i.e., $E(B-V)=0.50\pm0.46$ from the \citealt{greenred} map). 

\begin{figure}
\centering
\includegraphics[width=\linewidth]{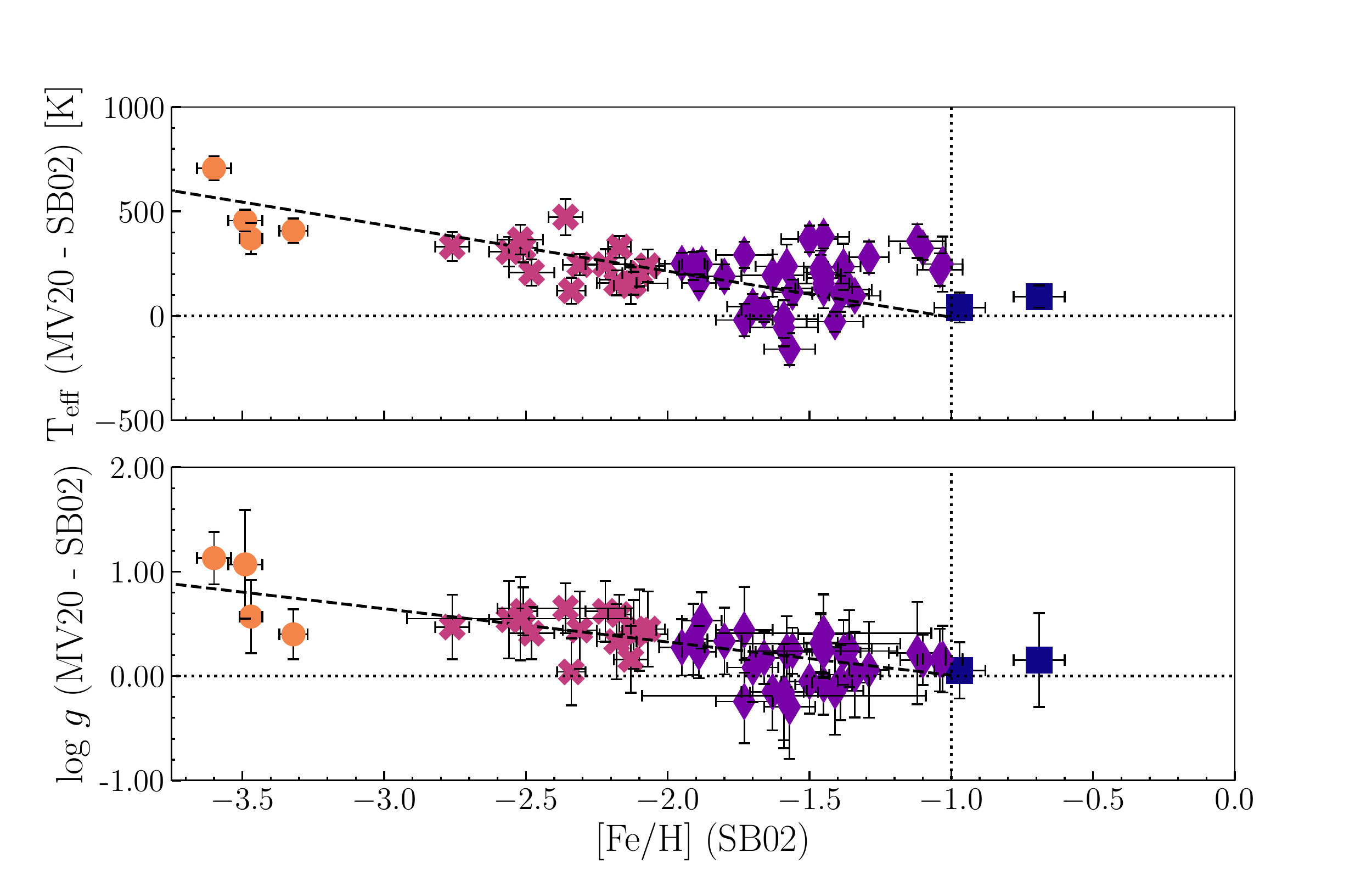}
\caption{\label{fig:mvsb} 
Offsets in stellar parameters (T$_{\rm eff}$ and log~$g$) with metallicity as determined from our \textit{isochrone-mapping} method compared with the spectroscopic results from SB02. Error bars from SB02 are included for each star. An offset of $\Delta$T${\rm eff}\sim+200$ K
and $\Delta$log~$g$~$\sim+0.3$ occurs per $\Delta\mathrm{[Fe/H]}=-1$ change in metallicity below $\mathrm{[Fe/H]}=-1$.}
\end{figure}

Average differences between our redetermined stellar parameters and those from SB02 are shown in Table \ref{tab:isooffset} for each bin, where this study is denoted as ``MV20''. The results show significant offsets and trends in the determination of stellar parameters as a function of metallicity.  
We find that the offsets in the stellar parameters determined from our \textit{isochrone-mapping} method scale as 
$\Delta$T$_{\rm eff}\sim+250$ K and
$\Delta$log~$g\sim+0.4$
per change of $\Delta\mathrm{[Fe/H]}=-1.0$ dex, below $\mathrm{[Fe/H]}=-1$;
see Fig.~\ref{fig:mvsb}.

\begin{table}
\caption{Average differences in effective temperature, surface gravity and metallicity between this study (``MV20'', see Appendix \ref{sec:spectromethod}) and that of SB02 (MV20-SB02) for each metallicity bin. The number of stars used to calculate each average is also included (N).}
\label{tab:isooffset}
\centering
\begin{tabular}{@{}lllll@{}}
\hline
Bin     & N & $<\Delta$T$_{\text{eff}} >$  & $<\Delta$log $g >$  & $<\Delta$[Fe/H] $>$ \\ \hline
[Fe/H] > -1.0  & 2  & $+55\pm38$   & $0.06\pm0.01$ & +0.02 \\
$-2.0<$[Fe/H]$\leq-1.0$ & 29    & $+179\pm119$	& $0.15\pm0.17$ & +0.05 \\ 
$-3.0<$[Fe/H]$\leq-2.0$ & 13 & $+440\pm104$ & $0.47\pm0.15$ & +0.20 \\
{[Fe/H]}$\leq-3.0$ & 4 & $+566\pm132$ & $0.79\pm0.30$ & +0.50 \\
\hline
\end{tabular}
\end{table}

\begin{table*}
\caption{Target information for our seven test case stars.  Star naming scheme from the Lowell Proper Motion Survey \citep{giclas71, giclas78}.
Two separate observations of G037-037 were made and co-added for the final SNR determinations.}
\label{tab:obs}
\begin{tabular}{llllllllllr}
\hline
Star	& Instrument Used	& Date Observed	& R. A.	& Decl.	& $V $ & $K$  &  Exposure	& S/N	 \\
	&				& (MJD)			& (J2000)	& (J2000)	&  &  & (s)			& (6500\AA)	\\
\hline
G184-007	& HIRES		& 51066.371 		& 18:24:13.099		& 27:17:10.896		& 14.42 &  12.33 & 3600 	& 115\\
G189-050	& HIRES		& 51067.459		& 22:56:27.490		& 33:53:04.200 	& 13.94 &  11.74 & 1800	& 180 \\
G158-100 & HIRES		& 51066.538		& 00:33:54.600		& -12:07:58.908	& 14.89 &  13.02 & 3600	& 135\\
G262-021	& HIRES		& 51066.429		& 20:35:25.560		& 64:54:04.716		& 13.94  &  11.74 & 3600	& 125 \\
G233-026	& GRACES	& 57242.618		& 22:39:56.351		& 61:43:07.561		& 11.98  &  10.03 &  3600	& 400 \\
G037-037	& GRACES	& 57373.329     & 03:23:38.352		& 33:58:30.310	& 11.89 &  10.71  & 1800  &  280 \\	
 ...        &  ...   & 57373.352	& ...	&
      ...      &  ... &  ... &  1800    &  ... \\
G241-004 & GRACES	& 57247.556		& 22:21:21.350		& 68:27:49.608		& 12.91 &  10.83 &  2250	& 20 \\
\hline
\end{tabular}
\end{table*}

The source of these offsets could be due to (1) uncertainties in the 1D~LTE model atmospheres analysis carried out by SB02, (2) the reddening estimates required in our \textit{isochrone-mapping} method, (3) assumptions made in generating the stellar isochrones  ([Fe/H], age, or alpha abundance), and/or (4) systematic errors in the isochrones and/or colour-temperature relations. Regarding the first point, standard 1D~LTE methods (as used by SB02) rely on high-quality model atmosphere models and radiative transfer analyses.  Improvements ranging from 3D to non-LTE effects \citep{amarsi16, chiavassa18, bergemann2012}, and in the atomic line lists \citep{denhartog2019, cowan2020}, have shown significant offsets for metal-poor stars, but not usually as large as those found here. The second and third points were investigated in Section \ref{sec:isochronemap}, where we found that uncertainties in reddening and age are the dominate sources of error for bins (i) and (ii).

For the metal-poor stars (in bins iii and iv), the difference in the stellar parameters between the isochrone-mapping method and spectroscopic method appears to be intrinsic and increasingly significant with lower metallicity.  A similar result has been seen for stars with metallicities [Fe/H] $< -2$, which can be modelled by exploring a range (of optimized values) for the convective mixing length parameter \citep{joyce15, joyce18}. Preliminary 1D~LTE analysis (as described in Section \ref{sec:spectromethod}) using the isochrone parameters, shows an increase of $\Delta$[Fe/H]=$+0.5$ for stars in the lowest metallicity bin (iv), a +0.2 dex for stars in bin (iii), and negligible increases for stars in bins (ii) and (i). To investigate this further, we re-derive the stellar abundances for all of the stars in bins (iii) and (iv) in Section \ref{sec:lowmetbins}, and compare the abundances derived using both spectroscopic and isochrone stellar parameters.

\subsection{Spectral Analysis of an SB02 Subset}

Significant improvements have been made in the field of stellar model atmospheres since SB02.  In particular, model atmospheres now include 1D, 3D, and <3D> radiative transfer, spherical extension, and overall improvements in our understanding of continuous and line opacities and broadening mechanisms.  Additionally, significant improvements have occurred in the precision of the atomic data (energy levels, oscillator strengths, hyperfine structure components, and NLTE line corrections).  Overall, these improve the precision in the absolute abundances of elements determined from the emergent stellar spectra. 

From the original SB02 sample, we selected a subset of stars for an updated detailed model atmospheres analysis, after applying additional selection criteria.  The additional selection criteria were as follows: \\

$\bullet$ Orbits with $R_{\text{apo}} > 20$ kpc, 

$\bullet$ Metallicities [Fe/H] $< -1.4$ dex, and

$\bullet$ Alpha-challenged with [$<$Mg,Si,Ca,Ti$>$/Fe]$<+0.2$.\\
 
The seven stars that satisfied these criteria are listed in Table \ref{tab:obs}. Note that the alpha abundances were determined as a weighted average of the available [$\alpha$/Fe] abundances by SB02. Four of the seven stars had existing Keck HIRES \citep{HIRES} spectra in the Keck archive, to study the remaining three we obtained new observations using the Gemini Remote Access to CFHT ESPaDOnS Spectrograph (GRACES) facility \citep{chen14}. A summary of the observation dates, target coordinates, magnitudes ($V$ from \cite{monet03}, $K_{s}$ from \cite{2MASSC}) and total spectral signal-to-noise ratio (SNR) for all seven stars is shown in Table \ref{tab:obs}. 

\begin{figure}
\centering
\includegraphics[width=\linewidth]{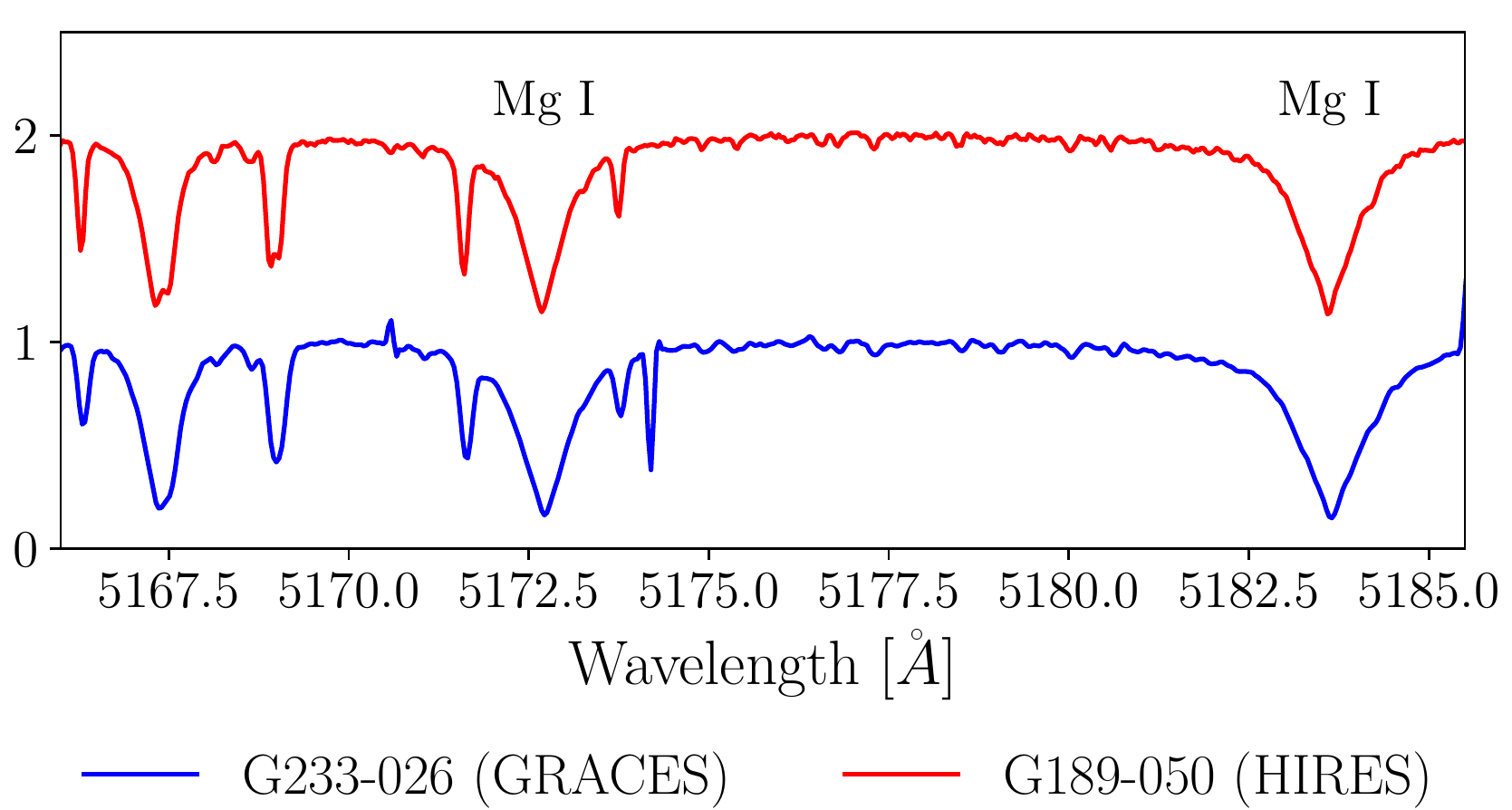}
\caption{Sample spectra near the Mgb lines is shown for G233-026 (\textit{blue}) from Gemini GRACES and for G189-050 (\textit{red}) from Keck HIRES.}
\label{fig:samplespec}
\end{figure}

\subsubsection{Gemini GRACES spectroscopy}
As mentioned, three of the SB02 stars in our sub-sample were observed with the Gemini GRACES spectrograph.  GRACES utilizes a 270-metre long optical fibre to combine the collecting power of the 8-metre diametre mirror of the Gemini North telescope with the exquisite spectral resolution of the Canada France Hawaii Telescope ESPaDOnS spectrograph \citep{espadons}. The GRACES fibre boasts a peak transmittance of 85\% at 800nm, with the ESPaDonS resolving power ranging from R = 40,000 to R = 66,000 across its 400 to 1,000 nm wavelength range. The GRACES spectra were collected in two fibre mode to yield better sky subtraction and observed over the course of three nights between August and December of 2015.  The resulting spectra displayed higher SNR at blue wavelengths (4500 \AA) than expected, given the long optical fibre coupling of GRACES.

\begin{table*}
\caption{Stellar Parameters determined from our classical LTE spectroscopic analysis are shown in the first line for each star, along with our GRACES/HIRES radial velocity measurements determined in \texttt{IRAF}. 
Stellar parameters determined from the \textit{isochrone-mapping} method described in Section \ref{sec:isochronemap} (combining \texttt{DSED} isochrones with Gaia DR2 parallaxes)  are listed in the second line. Also included in the second line are our NLTE corrected [Fe/H] values and the Gaia DR2 values of $T_{\mathrm{eff}}$ and radial velocity (in brackets, when available). All Gaia values of $T_{\mathrm{eff}}$ quoted have an associated uncertainty of 324K \citep{gaiateff}. Finally, the third line lists the original SB02 stellar parameters for each star. Values carried forward in our analysis are in \textbf{bold}.}
\label{tab:stellparam}
\begin{tabular}{llllllllr}
\hline
Star		& Method 	& $T_{\text{eff}}$			& $\log g$		& $M/M_{\odot}$		& [Fe/H]		& $\xi$			& $V_{\text{rad}}$ \\
			& 			& (K)						& (cm/$s^{2}$)	&					& 				& (km/s)	& (km/$s^{-1}$)	\\\hline
G037-037	& LTE 		& $6000\pm 200$	& $3.8\pm 0.3$				& ...				& $-2.35\pm0.14$			& $1.4\pm0.4$	& {$\mathbf{-143.0\pm 0.4}$} \\
        	& \texttt{DSED}\slash Gaia		& $\mathbf{6463\pm113}$ (5947)	& $\mathbf{4.41\pm0.04}$	& $\mathbf{0.74\pm0.04}$	 & $\mathbf{-1.96\pm0.19}$	& $\mathbf{1.4\pm0.1}$ 					& 
        	($-136.1\pm3.5$)\\ 
			& SB02 		& $5990\pm87$				& $3.76\pm0.24$	& ...				& $-2.36\pm0.06$			& $1.54\pm0.12$			& $-136.3\pm0.4$\medskip\\
G158-100  	& LTE 		& $5200\pm 200$				& $4.6\pm 0.1$	& ...				& $-2.36\pm0.13$			& $0.7\pm 0.5$			& $\mathbf{-360.6\pm 1.1}$  \\
        	& \texttt{DSED}\slash Gaia		& $(\mathbf{5346\pm57})$ (5321)	& $\mathbf{4.71\pm0.01}$	& $\mathbf{0.60\pm0.01}$	& $\mathbf{-2.24\pm0.15}$						& $\mathbf{0.6\pm0.1}$ 	
        	& \,\, (...) \\
			& SB02 		& $4981\pm71$				& $4.16\pm0.40$	& ...				& $-2.52\pm0.08$			& $0.50\pm0.32$			& $-357.9\pm 1.1$ \medskip\\
G184-007	& LTE 		& $5000\pm200$				& $4.5\pm0.2$	& ...				& $-1.77\pm0.14$	& $0\pm 0.5$	& $\mathbf{-370.6\pm 0.5}$\\
        	& \texttt{DSED}\slash Gaia		& $(\mathbf{5132\pm77})$ (5203)	& $\mathbf{4.71\pm0.01}$	& $\mathbf{0.58\pm0.01}$		& $\mathbf{-1.67\pm0.14}$		& $\mathbf{0.5\pm0.1}$ 					& \,\, (...) \\ 
			& SB02 		& $5147\pm90$				& $4.90\pm0.50$	& ...				& $-1.59\pm0.12$			& $0.0\pm0.5$			& $-371.7\pm 0.5$\medskip\\ 
G189-050	& LTE 		& $5400\pm 200$				& $4.5\pm 0.2$	& ...				& $-1.43\pm0.13$			& $0.2\pm 0.3$	& $\mathbf{-320.9\pm 0.6}$ \\
            & \texttt{DSED}\slash Gaia		& $(\mathbf{5463\pm110})$ (5412)	& $\mathbf{4.63\pm0.01}$	& $\mathbf{0.65\pm0.02}$	& $\mathbf{-1.41\pm0.15}$						& $\mathbf{0.8\pm0.1}$ 	
            & ($-322.0\pm1.2$)\\
			& SB02 		& $5254\pm82$				& $4.32\pm0.28$	& ...				& $-1.46\pm0.06$			& $0.3\pm0.3$ 			& $-320.7\pm 0.6$\medskip\\
G233-026	& LTE 		& $5400\pm300$		& $4.5\pm0.2$ 	& ...				& $-1.53\pm0.14$			& $1.1\pm0.3$	& $\mathbf{-318.63\pm 0.64}$ \\
        	& \texttt{DSED}\slash Gaia		& $(\mathbf{5503\pm62})$ (5473)	& $\mathbf{4.63\pm0.01}$	& $\mathbf{0.64\pm0.02}$	& $\mathbf{-1.34\pm0.15}$						& $\mathbf{0.8\pm0.1}$ 					& 
        	($-312.3\pm2.0$) \\ 
			& SB02 		& $5303\pm59$				& $4.39\pm0.26$	& ...				& $-1.45\pm0.06$			& $0.64\pm0.18$			& $-313.6\pm0.6$ \medskip\\
G262-021	& LTE 		& $5100\pm 300$	& $4.3\pm 0.2$	& ...				& $-1.37\pm0.13$			& $0.4\pm 0.4$	& $\mathbf{-214.0\pm 0.5}$\\
        	& \texttt{DSED}\slash Gaia		& $(\mathbf{5140\pm81})$ (5096)	& $\mathbf{4.67\pm0.01}$	& $\mathbf{0.60\pm0.01}$	& $\mathbf{-1.37\pm0.16}$						& $\mathbf{0.6\pm0.1}$ 					& 	\,\, (...) \\
			& SB02 		& $4985\pm56$				& $4.26\pm0.38$	& ...				& $-1.45\pm0.09$			& $0.00\pm0.50$			& $-214.5\pm 0.5$ \smallskip\\		
\hline
\end{tabular}
\end{table*}

The GRACES data were reduced using the IDL reduction pipeline {\sc DRAGRaces} ({\sc DR}) \citep{chen17} in two fibre mode using standard calibration images. The wavelength solution was calculated within {\sc DR} using the relevant ThAr arcs.  The 45th order could not be recovered for all three stars, and was left as a gap in the eventual 1D continuum-normalized spectra. The final output from {\sc DR} was a multi-extension \texttt{fits} file containing the recovered orders for both the science and sky fibres. As a final step in the reduction, the sky was subtracted and continuum normalized using k-sigma clipping; a nonlinear filter (a combination of a median and a boxcar) was used to smooth the mean pixels with an effective scale length for the filter set from 6 to 9 \AA, dependent on the crowding of the spectral lines.  We found that this was sufficient to follow the continuum without affecting the presence of the lines when used in conjunction with iterative ($\ge5$) clipping.
In the case G037-037, two independent spectra were stacked and the combination was re-normalized. Heliocentric corrections were applied following reduction using the \texttt{IRAF} \texttt{rvcorrect} task. Radial velocity corrections were performed using the {\sc IRAF} \texttt{fxcor} routine from lines in the proximity of the H$\alpha$ and the Mgb lines (near 518 nm); results are shown and compared with SB02's radial velocities in Table \ref{tab:stellparam}.  

Unfortunately, the SNR of the sky-subtracted spectrum for the star G241-004 was too low for further analysis.

\subsubsection{Keck HIRES (archival) spectroscopy}
Spectra for the remaining four stars in our sub-sample were retrieved from the Keck Observatory Archive\footnote{http://nexsci.caltech.edu/archives/koa/}. 
The spectra were taken with HIRES \citep{HIRES} during the SB02 observing campaign, which ran from July 1995 to  September 1998. SB02 noted that the spectrometer set up did not change appreciably from run to run; they selected the C1 decker to define the slit dimensions as 0.861" wide by 7.0" long and used a KV408 order blocking filter to eliminate contamination from neighboring diffraction orders.  The HIRES red collimator/camera and the Tek 2048 CCD were used to gather spectra from 450 to 680 nm, with small inter-order gaps redward of 500 nm. Similar to SB02, we used standard \texttt{IRAF} reduction methods as in \texttt{noao.imred.echelle}. Science images were de-biased, flattened and trimmed to account for the overscan regions. Sky subtraction was performed during the 1D extraction and bad pixels were identified and removed.  Wavelength solutions were created using the corresponding ThAr spectra and used to calibrate the final 1D spectra. These 1D spectra were continuum normalized using k-sigma clipping (as described above for the GRACES spectra). Radial velocity corrections were applied using the \texttt{IRAF} \texttt{dopcorr} task and heliocentric corrections using the same method as the GRACES data.  

\begin{figure*}
\centering
\includegraphics[scale=0.6]{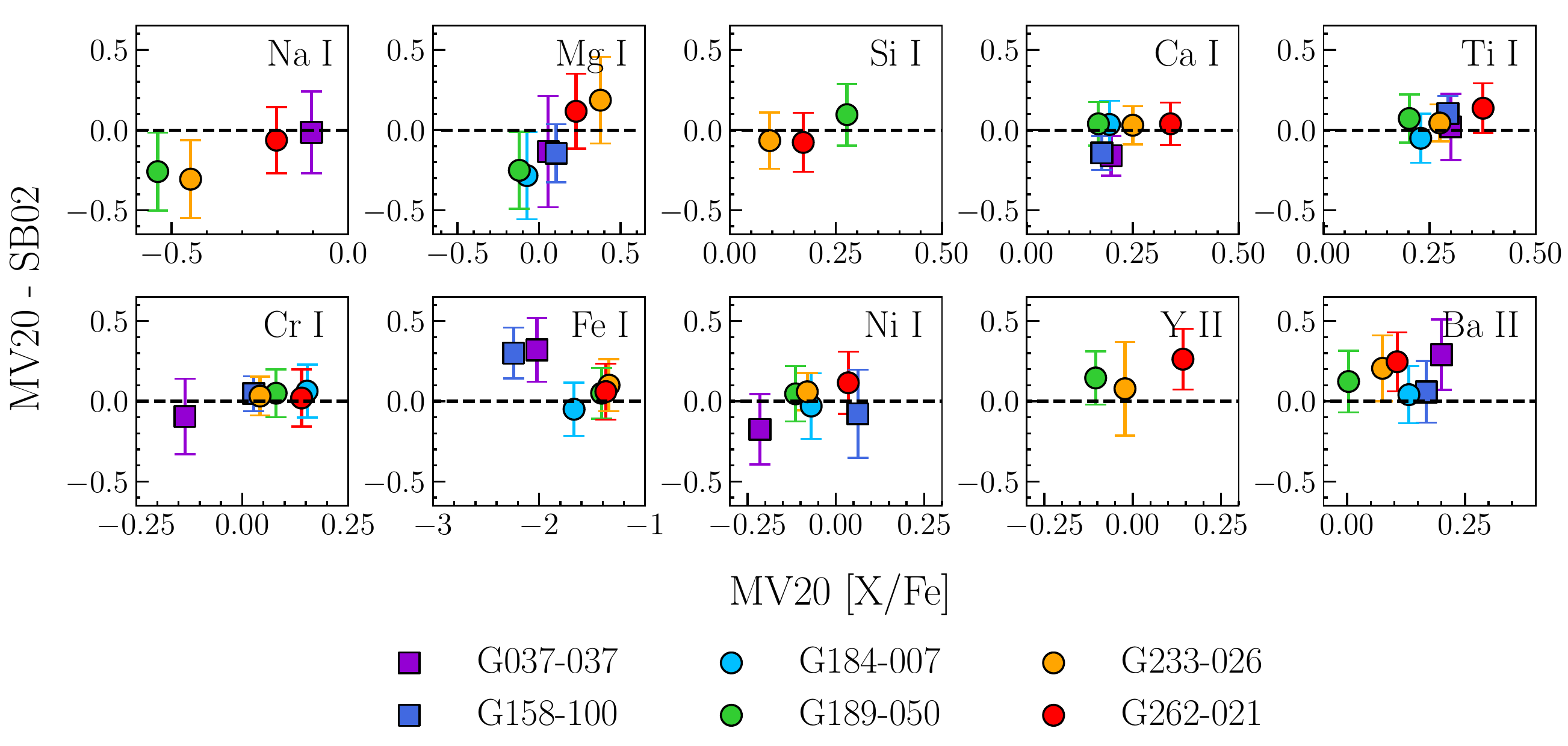}
\caption{\label{fig:sb02mv_offset} A comparison of the element abundances [X/Fe] between our analysis ``MV20" and SB02 for the six stars in our sub-sample.  The x-axis is [X/Fe], except for \ion{Fe}{I} where [\ion{Fe}{I}/H] is used. Abundances have been scaled to the \citet{asp09} solar abundance scale, and are listed in Table \ref{tab:abund}. Stars that do not appear in all ten plots were missing elemental abundances in one or both of the studies. The lowest metallicity stars in our subset found in bin (iii) are shown using square markers.}

\includegraphics[scale=0.58]{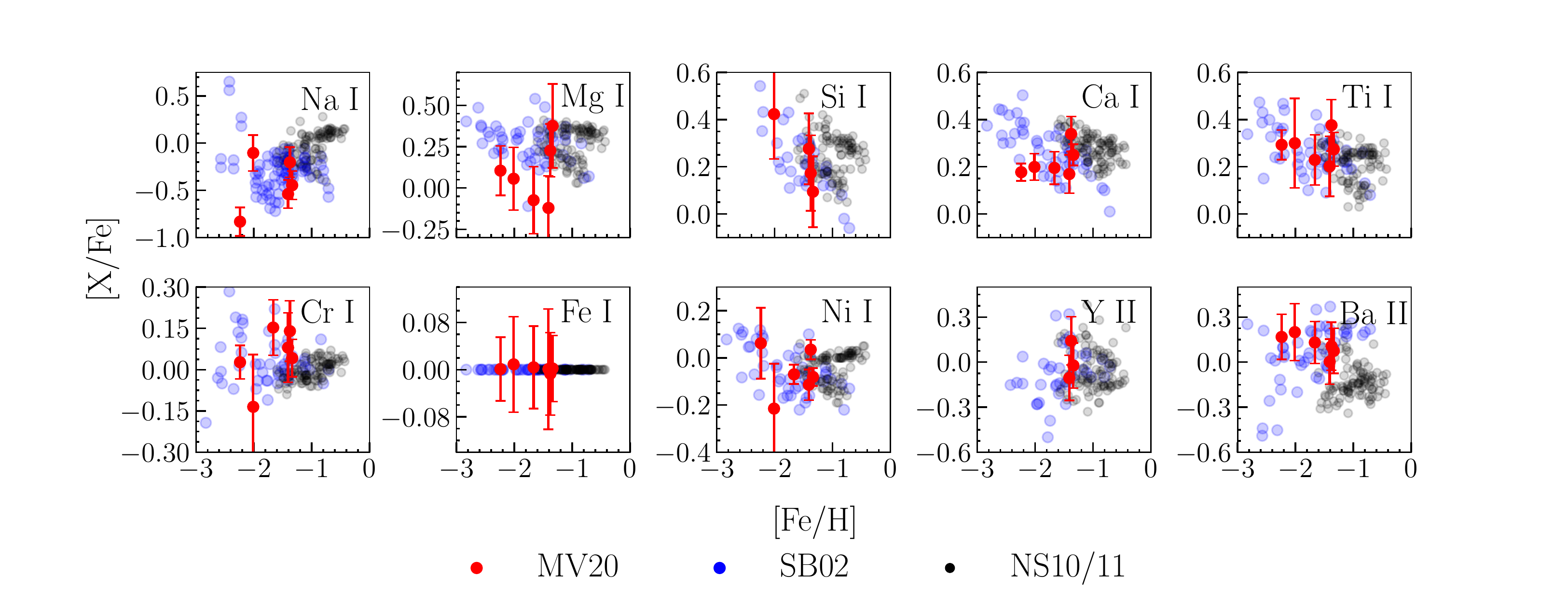}
\caption{\label{fig:sb02mv_all} Comparison of the abundances from SB02 (blue), NS10/11 (black), and our analysis of the six stars in our subset ``MV20" (red) for 10 elements in common.
Updated abundances are used for the six stars in our sub-sample and the stars in bins (iii) and (iv), original abundances are adopted for stars in bins (i) and (ii).}
\end{figure*}

\section{Stellar Abundances}\label{sec:stellabund}

\subsection{Subset of six stars with GRACES or HIRES spectra}

For our subset of six stars with GRACES or HIRES spectra, the stellar parameters are determined using both the \textit{isochrone-mapping} method and an independent classical analysis using the \ion{Fe}{I} and \ion{Fe}{II} spectral lines (see Appendix \ref{sec:specparams}).Our results for both sets of stellar parameters are shown in Table \ref{tab:stellparam}, along with those from SB02.   
 
The chemical abundances are determined for each set of stellar parameters from classical model atmosphere analysis. Model atmospheres were generated using both \texttt{MARCS} \citep[][with additions by B. Plez]{MARCS1, MARCS2} and ATLAS models \citep{atlas12}.  Spectral lines and atomic data were compiled from SB02, \cite{nor17} and \cite{bat17}, after limiting the range of exploration to the overlapping wavelength regimes of the HIRES and GRACES data ($4200 - 6700$ \AA).  Atomic data were updated when appropriate by comparing to the {\sl linemake}\footnote{{\sl linemake} contains laboratory atomic data (transition probabilities, hyperfine and isotopic substructures) published by the Wisconsin Atomic Physics and the Old Dominion Molecular Physics groups.  These lists and accompanying line list assembly software have been developed by C. Sneden and are curated by V. Placco at https://github.com/vmplacco/linemake.} atomic and molecular line database.  Isotopic and hyperfine structure corrections were also examined (for lines of \ion{Ba}{II}, \ion{Mn}{II}, and \ion{Li}{I}), but no significant corrections were found (all targets are dwarf stars). 
Equivalent widths (EWs) were measured as described in Appendix \ref{sec:ewmeas} and used in the 1D~LTE radiative transfer code MOOG \citep{moog} to determine the chemical abundances.

Chemical abundances are compared to the Sun using the  \cite{asp09} solar data, and listed in Table \ref{tab:abund}. Final abundance errors were calculated from the line-to-line abundance dispersion ($\sigma_{\text{EW}}$) added in quadrature with the uncertainties imposed by the stellar parameter errors
($\sigma_{Teff}$, $\sigma_{\text{log}g}$, $\sigma_{\text{[Fe/H]}}$, and $\sigma_{\xi}$).

\subsection{\label{sec:sb02} Comparisons with SB02}

From the detailed analyses of six stars in our subset (those with GRACES or HIRES spectra), we find excellent agreement with SB02 for the four stars with [Fe/H] $> -2$ (G184-007, G189-050, G233-026, G262-021; e.g., see Table \ref{tab:stellparam}).  
All stellar parameters (from \textit{both} of our stellar parameter determination methods) and SB02 are within 1$\sigma$ errors.
Chemical abundance differences between this analysis (MV20) and SB02 are shown in Fig.~\ref{fig:sb02mv_offset}. 
Again, there is excellent agreement for most of the chemical abundances;   
exceptions are the abundances of \ion{Na}{I} (G189-050, G233-026), \ion{Mg}{I} (G189-050, G184-007), \ion{Y}{II} (G262-021 and G189-050) and \ion{Ba}{II} (G262-021, G233-026).   The differences in Na and Mg are discussed in subsequent sections, while the differences in \ion{Y}{II} and \ion{Ba}{II} are attributed to heightened sensitivities to small differences in the stellar parameters.
\textit{We note that there is excellent agreement with SB02 when their original spectroscopic stellar parameters are adopted (once adjusting to the same solar abundances scale and atomic line list).   
}

The remaining two stars (G037-037, G158-100) have metallicities [Fe/H] $\leq -2$, where we find significant differences in the stellar parameter results from our spectroscopic\footnote{We note that these differences could not be attributed to neglected NLTE effects on the \ion{Fe}{I} lines; see Appendix \ref{sec:NLTE}.}  and \textit{isochrone-mapping} methods.   These two stars are in bin (iii), as discussed in Section \ref{sec:sb02gaia} (also see Table \ref{tab:isooffset}), where significant offsets are found throughout our reanalysis of the SB02 sample. The \ion{Fe}{I} difference between MV20 and SB02 in these two stars is a consequence of the large offsets in log $g$ ($\sim0.6$) and temperature ($\sim400$ K). However, the difference in [Fe/H] does not result in disagreements larger than $1\sigma$ in any of the remaining \textit{relative} abundance ratios [X/Fe] (see the square markers in Fig.~\ref{fig:sb02mv_offset}). Therefore, for the sake of consistency regarding the treatment of our small sub-sample, we choose to adopt the stellar parameters associated with the \textit{isochrone-mapping} method for these stars. 

Given the disagreement between the isochrone and spectroscopically-derived abundances and the observed failure of the low-metallicity isochrones to reproduce the luminosities for the majority of stars with [Fe/H]$<-2$ (discussed in Section \ref{sec:lowmetbins}), we chose to adopt the original SB02 stellar parameters for the remaining stars in bins (iii) and (iv).  Hence, using the SB02 parameters, we have updated the abundances for the lowest metallicity stars.  In general, the chemical abundance ratios [X/Fe] remain unchanged from SB02, with a few exceptions;  those include \ion{Ca}{I}, \ion{Ba}{II}, and \ion{Ti}{I} in some stars, where our new [Ca/Fe], [Ba/Fe], and [Ti/Fe] ratios are offset by $\sim-0.1$, $\sim+0.2$, and $\sim+0.05$, respectively (see Fig.\ref{fig:updatedabund}). \textit{These new abundances generated using the SB02 stellar parameters, with our updated atomic data and model atmospheres, are adopted for the bin (iii) and bin (iv) stars throughout the rest of this paper.}

\begin{figure}
\centering
\includegraphics[width=\linewidth]{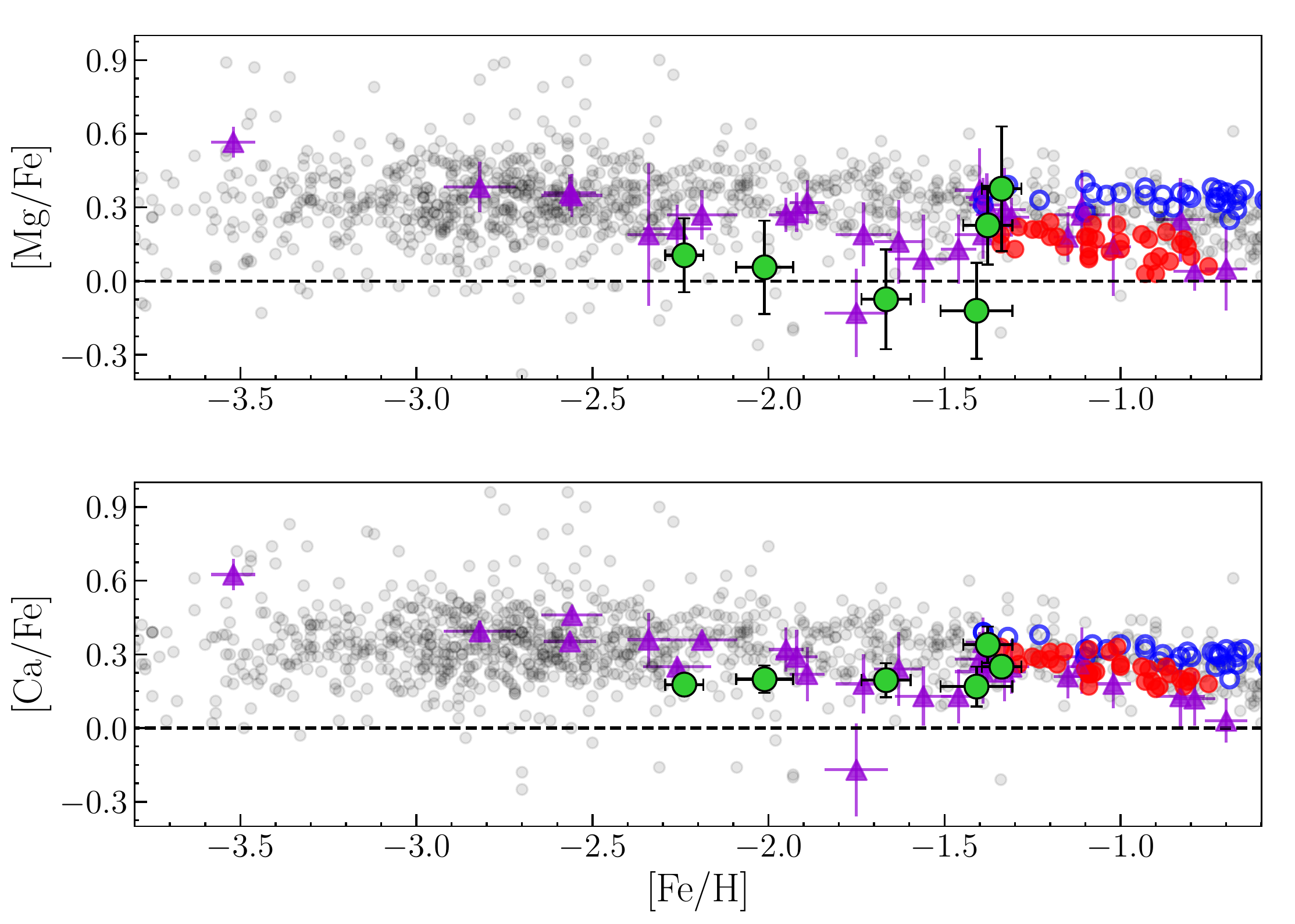}
\caption{\label{fig:sb02ns1011_mgca} Alpha abundances for our ab initio analysis of six SB02 stars (green), and the entire SB02 sample (purple triangles, including our reanalysis for the metal-poor stars; see text), as well as the low alpha (red) and high alpha (blue) stars from NS10/11 \citep{niss10, niss11}, and MW halo stars (grey; \citealt{yong2013}, \citealt{berg15}, \citealt{venn04}).}
\end{figure}

\begin{figure}
\centering
\includegraphics[width=\linewidth]{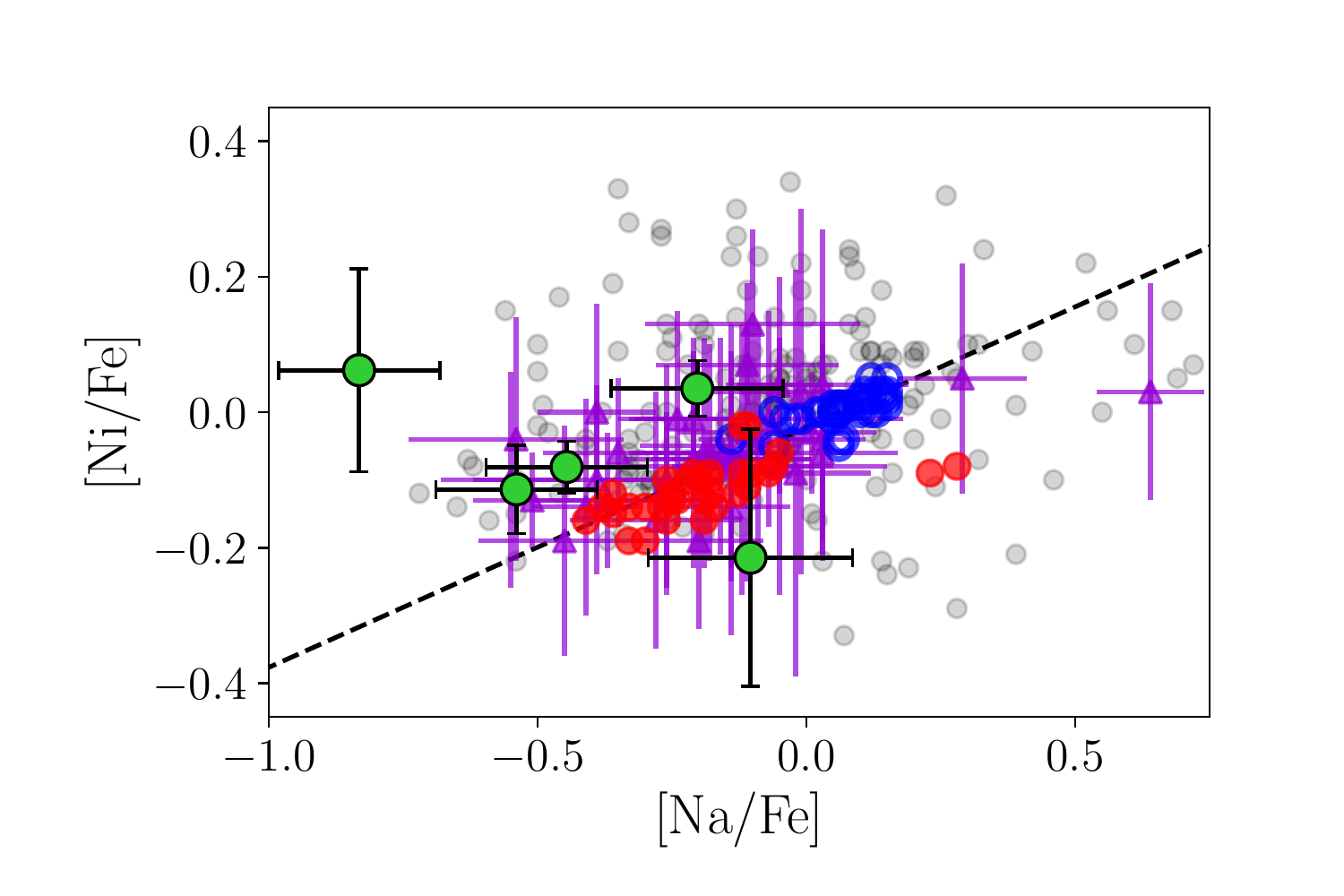}
\caption{\label{fig:sb02ns1011_nani} Na-Ni correlation found by NS10/11 for the three studies, using the same labels as in Fig.~\ref{fig:sb02ns1011_mgca}. We note that the three points from this paper (green) that are closest to the Ni-Na line are also Mg-poor stars. NLTE corrections have been applied to the Na abundances for the stars in this study, NLTE corrections to Na were not applied by SB02 or NS10/11.}
\end{figure}

\subsection{\label{sec:comp_ns1011}Comparisons with NS10/11}

We now compare the analysis of our subset of six stars with the high-velocity, intermediate metallicity ([Fe/H]$>-1.6$) stars examined by \cite{niss10, niss11}. 
In Fig.~\ref{fig:sb02mv_all}, the NS10/11 sample are compared with the original SB02 sample, and the results from our detailed spectral analysis of the subset of six stars. 
All abundances have been scaled to the \citet{asp09} solar abundances, however we could not shift the NS10/11 data due to their differential abundance methodology. 

Excellent agreement is found between the majority of stars over a range of metallicities in these three data sets. No systematic offsets are seen between these three studies, with the exception of \ion{Mg}{I}, \ion{Ba}{II}, and one star in \ion{Na}{I} (G158-100). Regarding \ion{Mg}{I}, our abundances are in good agreement with SB02; however, we purposely selected metal-poor, alpha-challenged stars for our subset (see Section \ref{sec:sb02gaia}) and therefore expect that our six stars are truly low in [Mg/Fe]. Regarding \ion{Ba}{II}, again our abundances are in good agreement with SB02 (when the same stellar parameters are adopted); however, both of these studies show an offset $\sim$+0.2 dex above the NS10/11 sample, over the entire metallicity range. The source of this offset is due to a difference in the atomic lines used; both analyses use the weak \ion{Ba}{II} 5854 and 6141 \AA\ lines, but SB02 also use the stronger 4554 \AA\ and the slightly blended 4934 \AA\ features. Regarding \ion{Na}{I}, SB02 did not measure Na in G158-100 as they rejected lines with an EW $\ge$ 85 \AA; we measure and use \ion{Na}{I} $\sim5895$ \AA \, at EW $\sim150$ \AA; see Section \ref{sec:spectromethod}).  

Alpha abundances ([Mg/Fe] and [Ca/Fe]) are examined in more detail relative to the NS10/11 sample in Fig.~\ref{fig:sb02ns1011_mgca}.  NS10/11 found two distinctly different groups of stars based on a differential abundance analysis and detailed kinematics: stars with high alpha abundances and halo kinematics (blue), and alpha-poor stars with high or retrograde velocity orbits (red). Not only are the majority of our six stars and the rest of the SB02 sample more metal-poor than NS10/11, but our Mg-poor stars do not fit with either of the NS10/11 groups.  If the low-alpha stars in the NS10/11 sample have been accreted from a dwarf galaxy, our four Mg-poor stars from the SB02 sample are from a different accretion event (or events). 

NS10/11 also found that the [Ni/Fe] and [Na/Fe] abundances were correlated, but offset, between their two groups; see Fig. \ref{fig:sb02ns1011_nani}. The Ni and Na abundances in the SB02 sample are in good agreement with the NS10/11 sample.  From our ab initio analyses of our six stars, there is an offset from the Ni-Na relationship towards higher [Ni/Fe] (or equivalently lower [Na/Fe]) values for four stars. This offset may be due to the large NLTE corrections applied to our sample, e.g., NLTE corrections were as large as $-0.4$ dex for the \ion{Na}{D}~5895\AA~line in the two lowest metallicity stars. The NS10/11 line list is not publicly available, however it is clear that they did not apply NLTE corrections. Modulo the NLTE corrections, we find that our three (Mg-poor) stars are in good agreement with the NS10/11 $\alpha$-poor stars that define the lower part of their Na-Ni trend.

\begin{figure}
\centering
\includegraphics[width=\linewidth]{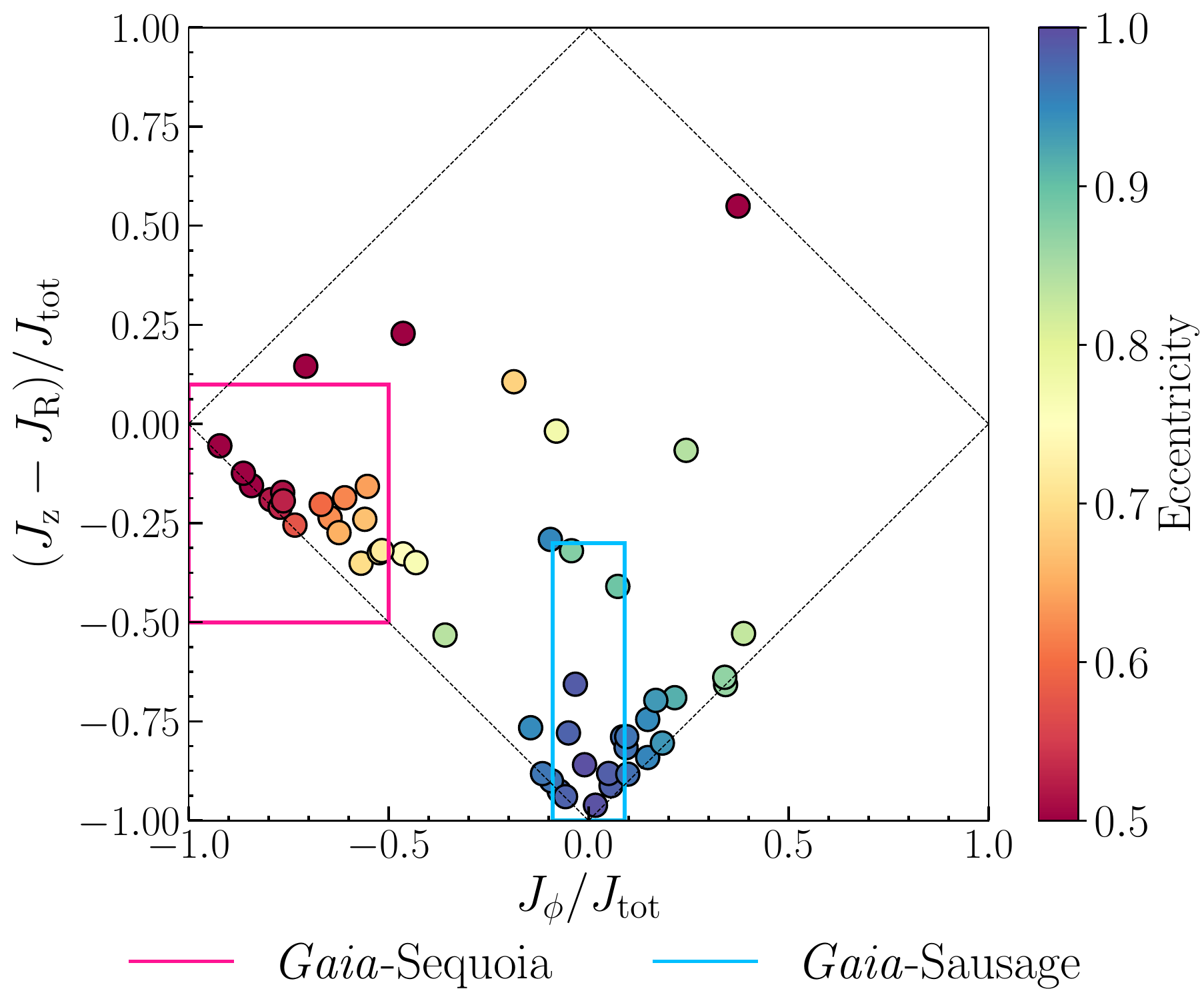}
\caption{Action map of the entire SB02 sample. The horizontal axis shows the $J_{\phi}$ action, and the vertical axis shows the difference between the vertical and radial action, with both axes normalized by the total action. The approximate locations of the \textit{Gaia}-Sausage (blue box) and \textit{Gaia}-Sequoia (pink box) events are identified as determined by \citep{myeong19}}
\label{fig:actionmap}
\end{figure}

\section {Stellar Dynamics}{\label{sec:orb}}

It has been demonstrated that the best way to identify stars belonging to discrete merger events, such as \textit{Gaia}-Sausage or \textit{Gaia}-Sequoia, is by using their dynamical properties, i.e., their orbital actions retain the signature of the event over long time periods \citep{myeong18a,myeong18b,myeong19}. Here we calculate dynamical properties of the SB02 sample from Gaia DR2 data (see Appendix~\ref{sec:kinematicdata})
to investigate potential associations with ancient merger events.  

To determine the dynamical properties, we adopt the potential in \citet{mcmillan17}, the characteristics of which are described in Tables 1 and 4 of that work. Celestial coordinates are transformed to galactocentric coordinates assuming the following: the location of the galactic center is (RA = 17:45:37.224 h:m:s, Dec = -28:56:10.23 degrees) \citep{reid04}; The Sun is 20.8~pc above the galactic plane \citep{bennett19} and the solar peculiar velocity is $(U,V,W)=(11.1,12.24,7.25)$~km/s \citep{schonrich10}. The distance to the solar circle, and the circular velocity at that radius are $8.121$~kpc \citep{gravity18} and $229$~km/s \citep{eilers19} respectively, which are both quite similar to the intrinsic values of the \citet{mcmillan17} potential. Note that we use a left handed coordinate system such that, from the position of The Sun, galactocentric X, Y, and Z are positive towards galactic anti-center, the direction of galactic rotation, and the galactic north pole respectively. Orbital integrations were performed using a 4-D symplectic integrator implemented in \texttt{GALPY}, a galactic dynamics Python package \citep{bovy15}. All orbits were integrated for a total of 10 Gyr ($\pm5$ Gyr) with select orbits shown in Appendix \ref{sec:kinematicdata}.

\subsection{\label{sec:action} \textit{Gaia}-Sausage and \textit{Gaia}-Sequoia}
The actions $J_{\phi}$, $J_{\text{r}}$ and $J_{z}$ of the SB02 sample were calculated using an implementation of the St$\ddot{\text{a}}$ckel fudge method in \texttt{GALPY} \citep{bin12, mack18}. Fig.~\ref{fig:actionmap} shows our sample in action space, where the horizontal axis shows the $\phi$ action (equivalent to $L_{z}$) and the vertical axis shows the difference between the vertical and radial actions. Both axes are normalized by the absolute sum of the actions: $J_{tot}$. We label the \textit{Gaia}-Sausage and \textit{Gaia}-Sequoia accretion events following the convention of \cite{myeong19}\footnote{Note that the boundary of the \textit{Gaia}-Sausage stars has been extended to include stars with $|J_{\phi}/J_{\mathrm{tot}}|<0.09$ following the suggestion of G.C. Myeong through private communication.}. While it broadly appears that our sample is dominated by stars populating these two regions, the selection criteria that define our sample undoubtedly plays a role in sculpting the appearance of the distribution. Each star in Fig.~\ref{fig:actionmap} is coloured by its eccentricity, and in general we see that stars in the \textit{Gaia}-Sausage selection box have $e > 0.9$ and those in the \textit{Gaia}-Sequoia selection box have $e \sim 0.5--0.6$, which is as expected.

The Gaia data for the SB02 sample was filtered to remove stars with $BP-RP$ ($>2$) and \texttt{phot\_bp\_rp\_excess} ($\sim>1.4$) in Fig.~\ref{fig:actionmap}.  Stars with large \texttt{astrometric\_chi2\_al} values, indicating a poor astrometric solutions, were also removed. These cuts follow from \cite{gaiacalib}, who recommend astrometric solutions with large \texttt{astrometric\_chi2\_al} values be avoided, while values of \texttt{phot\_bp\_rp\_excess\_factor} should be around one for normal stars. Interestingly, the star that displays the most circular orbit (shown as the star closest to the top vertex), G025-024, has normal astrometric parameters, suggesting that is in-fact unique dynamically from the remaining stars in the sample.

In our subset of six SB02 stars, G262-021 and G233-026 are classified as belonging to the \textit{Gaia}-Sausage accretion event, while G158-100 and G184-007 are identified as belonging to the \textit{Gaia}-Sequoia event.  For the remaining stars in the SB02 sample, we only identify stars found entirely within the action-space bounds as possible members of each accretion event. In total 11 SB02 stars are classified as \textit{Gaia}-Sausage stars, and 17 are classified as \textit{Gaia}-Sequoia stars.

\begin{figure*}
    \centering
    \includegraphics[scale=0.6]{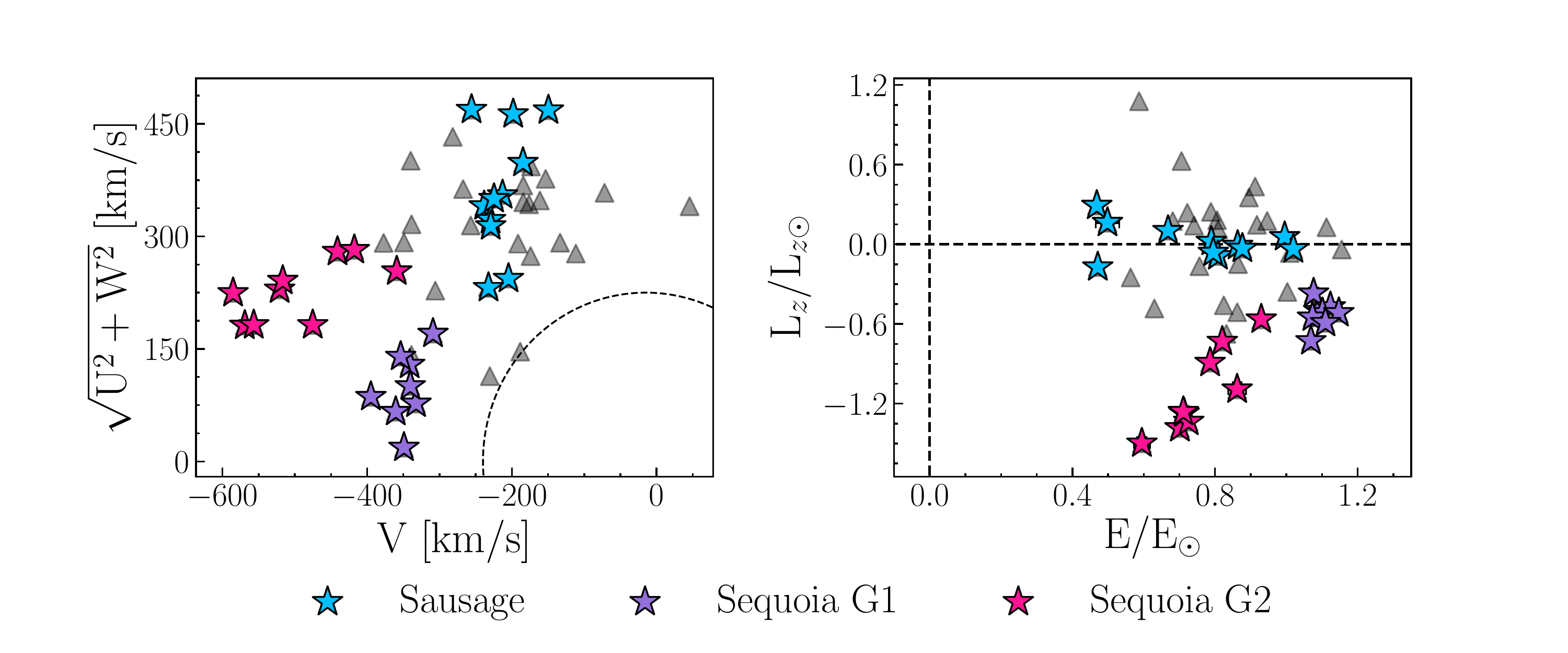}
    \caption{Toomre diagram (\textit{left}) showing stars in the SB02 sample.  Those associated with the \textit{Gaia}-Sausage merger (sky blue stars), the two dynamical subsets of the \textit{Gaia}-Sequoia event (pink stars and purple stars), or simple outer halo stars (grey) are identified.  Angular momentum and energy (\textit{right}) are scaled by the \citep{mcmillan17} potential solar values ($L_{z\odot}=2014.2$ kpc km/s and $E_{\odot} = -1.54\times10^{5}$ km$^{2}$/s$^{2}$).  The dashed line represents stars with thin disk dynamics (V$_{\rm circ}$ = 229 km$s^{-1}$).}
    \label{fig:toomre}
\end{figure*}

\begin{figure}
    \centering
    \includegraphics[width=\linewidth]{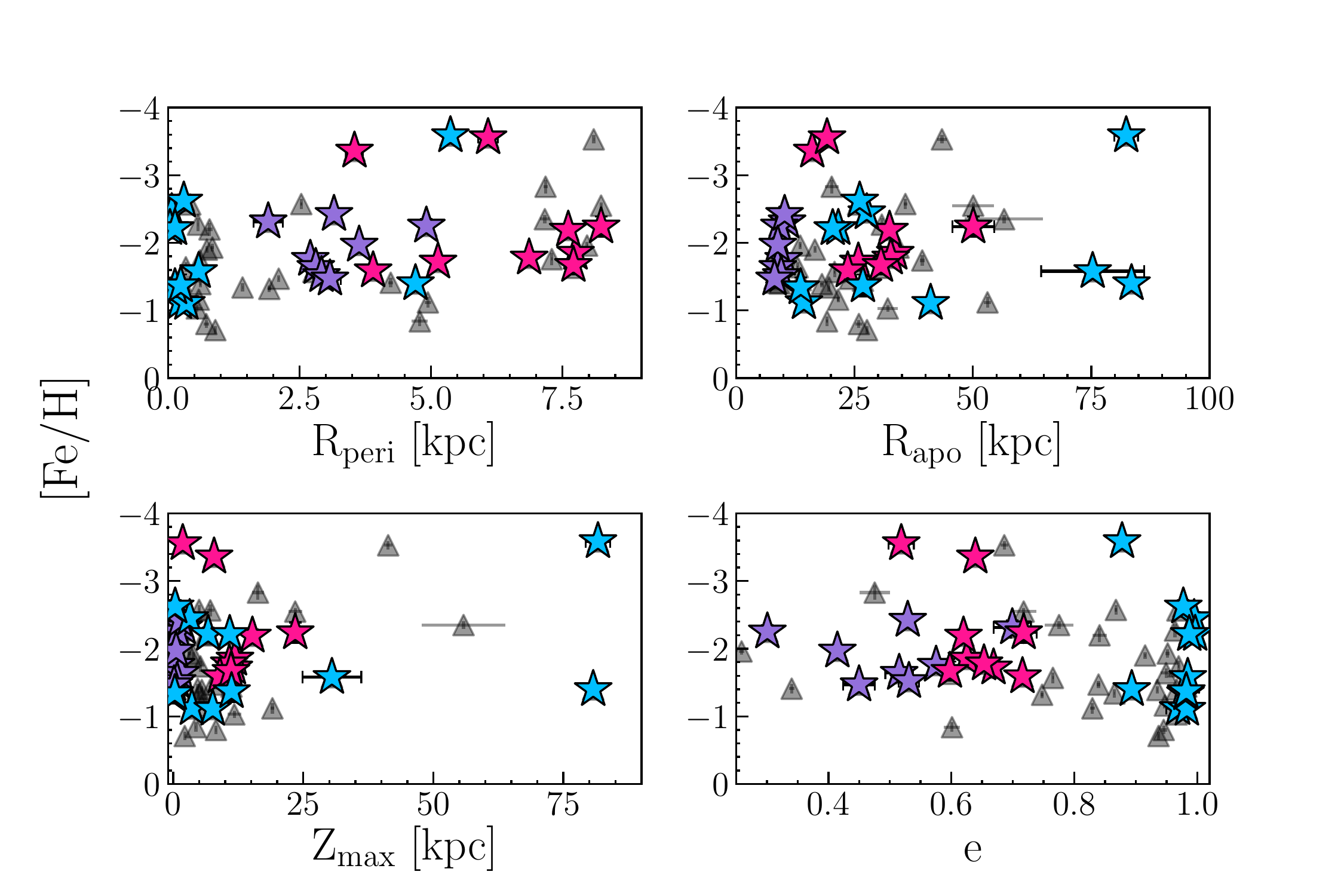}
    \caption{Metallicity distribution as a function of various orbital parameters derived in Section \ref{sec:orb}. Data is presented using the convention of Fig.~\ref{fig:toomre}.}
    \label{fig:mgfehgeseq}
\end{figure}
 
\subsection{\label{sec:chemodyn} Chemo-dynamics of the SB02 sample}
The distribution of the SB02 sample is shown on the Toomre diagram and in angular momentum vs. energy space ($L_{z}$ vs. $E$) in Fig.~\ref{fig:toomre}.  Values of $L_{z}$ and $E$ shown are normalized by the solar values from the \cite{mcmillan17} potential ($L_{z\odot}=2014.2$ kpc km/s and $E_{\odot} = -1.54\times10^{5}$ km$^{2}$/s$^{2}$.  In the Toomre diagram, it is clear that the majority of the SB02 sample are high velocity halo stars or on highly retrograde orbits, confirming the initial kinematic cuts made by SB02. A clear separation between the \textit{Gaia}-Sausage and \textit{Gaia}-Sequoia stars is also seen, as is expected given their unique dynamical signatures. In $L_{z}$ vs $E$ space the \textit{Gaia}-Sequoia stars are more distinct with highly retrograde orbits, while \textit{Gaia}-Sausage stars straddle the line of zero angular momentum, which reflects the radially biased orbits of its constituent stars. While we employ a different potential than \citet{helmige}, the broad trends which describe the kinematics of these merger events remain. 

Interestingly, we find two distinct dynamical subsets within the \textit{Gaia}-Sequoia stars, visible in both the Toomre and in $L_{z}$ vs $E$ space. We have formalised this distinction by splitting the \textit{Gaia}-Sequoia stars into a low orbital energy group, those with $E/E_{Mdot}\geq1.0$ (G1) and a high orbital energy group, those with $E/E_{Mdot} < 1.0$ (G2). This distinction seems to be in agreement with the findings of \cite{myeong19}, where the two Sequoia sub-groups could explain the extended regions associated with \textit{Gaia}-Sequoia in action space. \cite{yuan2019} also find two independent retrograde groups associated with \textit{Gaia}-Sequoia (``DTG-4'' and ``DTG-5''), both of which have clearly distinct mean energies. Finally, \cite{koppleman2019} also find a natural division of \textit{Gaia}-Sequoia into high and low orbital energy groups, however they attribute the division as evidence that Sequoia stars do not originate from a single progenitor. Instead they assign the high orbital energy stars as true \textit{Gaia}-Sequoia stars and the low orbital energy stars as belonging to a separate accretion event, \textit{Gaia}-Thamnos. \cite{koppleman2019} also show a chemical distinction between the two sub-groups which they further attribute as evidence that they are associated with different accretion events. To investigate this further within our own sample we carry on the distinction between the two \textit{Gaia}-Sequoia groups, assessing them as potentially unique events.
    
The distribution of metallicity is examined as a function of several orbital parameters in Fig.~\ref{fig:mgfehgeseq}, including pericentric radius (R$_{\text{peri}}$), apocentric radius (R$_{\text{apo}}$), maximum height from the disk (Z$_{\text{max}}$) and eccentricity ($e$).  While both events show unique dynamic signatures, no obvious chemo-dynamic trends are found in either of the major groups, nor in the \textit{Gaia}-Sequoia sub-groups. A chemo-dynamic trend if present could probe the hypothesis that, the outer parts of the accreted systems were stripped first, then the location of associated member stars within the MW potential could probe the existence of an original metallicity gradient in the progenitor.  Either these systems were small enough that no metallicity gradient existed, or this sample is too small to test this hypothesis.

\subsubsection{\label{sec:chemsig}Chemical signatures of accretion: locating the [$\alpha$/Fe] knee}
 
\begin{figure}
    \centering
    \includegraphics[width=\linewidth]{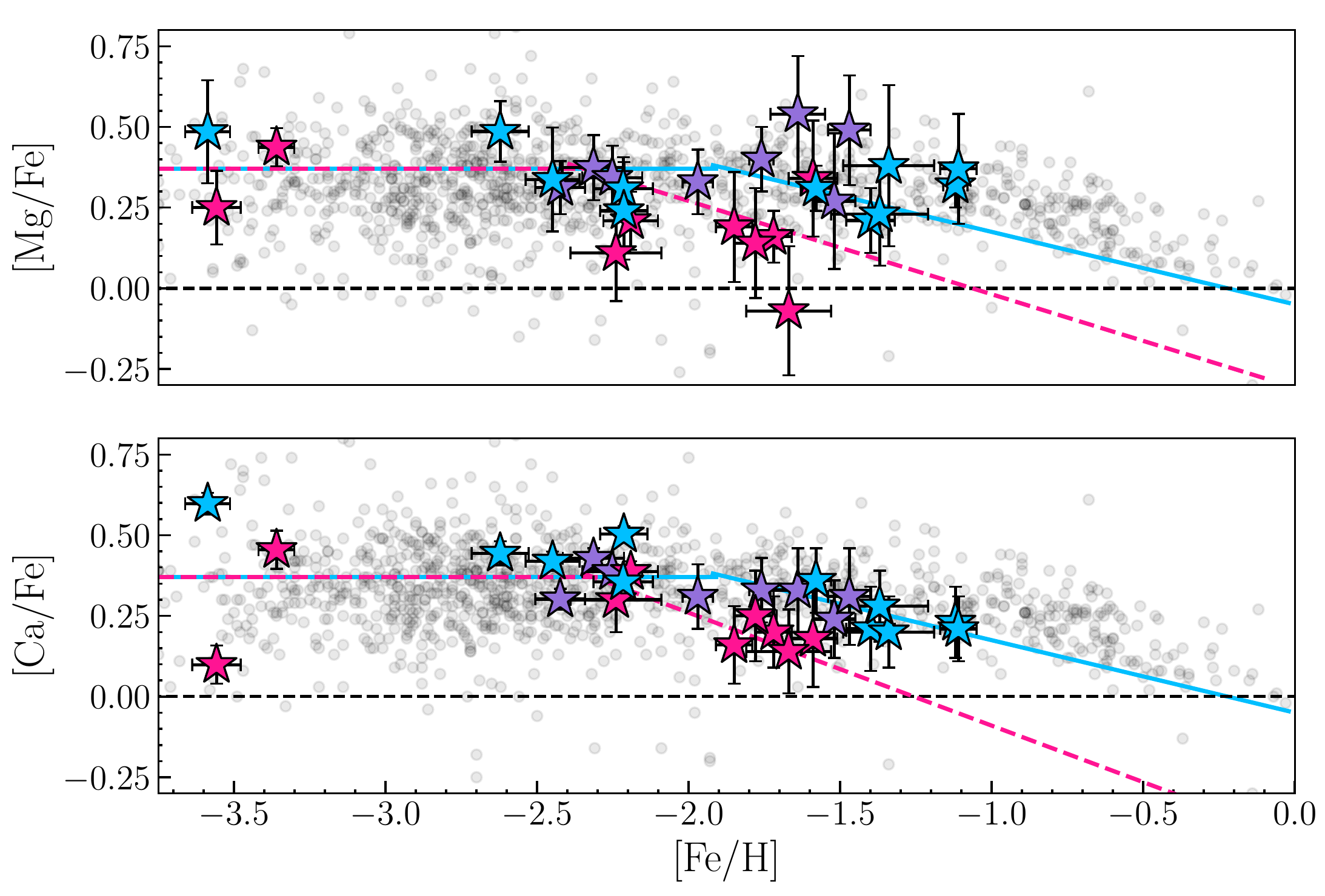}
    \caption{Alpha abundances for stars classified as part of the proposed \textit{Gaia}-Sausage accretion event (sky blue stars) and two sub-groups of the \textit{Gaia}-Sequoia accretion event (purple and pink stars). The location of the alpha knee in both the \textit{Gaia}-Sausage stars and the high orbital energy (G2) \textit{Gaia}-Sequoia stars is plotted as the decreasing trends around [Fe/H]$\sim-1.6$ (sky blue line) and [Fe/H]$\sim-2.3$ (pink dashed line) respectively. Halo stars from the literature are also included as black circles \citep{yong2013, berg15, venn04}. }
    \label{fig:capturedmgca}
\end{figure}

One of the classic indicators of the star formation history in a dwarf galaxy is the metallicity of the [$\alpha$/Fe] knee, i.e., where [$\alpha$/Fe] begins to decrease as a function of increasing metallicity \citep{tolstoy09, venn04}.  The knee is usually attributed to the onset of Type Ia supernovae, diluting the alpha abundances produced from earlier core collapse supernovae, although it may also be related to variations in the local IMF \citep{tinsley79, matt90, matt03, tolstoy03, mcwilliam13, fernandez2018}.  A slower star formation rate, or effectively truncated upper IMF, moves the knee to lower metallicities in dwarf galaxies. Assuming that a large fraction of the SB02 stars (4/6 in our sub-sample and 28/54 in the entire re-analyzed sample) may be associated with one or more proposed merger events, we examine the [Mg/Fe] and [Ca/Fe] ratios vs [Fe/H] in Fig.~\ref{fig:capturedmgca}.

For the stars we associate with \textit{Gaia}-Sausage, an [$\alpha$/Fe] knee appears in both Mg and Ca at lower metallicities than the Galactic comparison stars, near [Fe/H]$\sim-1.6$; see Fig.~\ref{fig:capturedmgca}. This value is in good agreement with the NS10/11 low-alpha stars ([Fe/H]$\leq-1.5$), slightly lower than the location identified by \cite{myeong19} using SDSS APOGEE data ([Fe/H]$=-1.3$), and slightly higher than \cite{matsuno19} using the SAGA database ([Fe/H]$\sim-2$).

For the stars that we associate with \textit{Gaia}-Sequoia, an [$\alpha$/Fe] knee is much less distinct if the two groups are treated as one, but visible if the groups are examined separately. The \textit{Gaia}-Sequoia G1 stars have an [$\alpha$/Fe] knee nearly coincident with that of the \textit{Gaia}-Sausage stars. The location of the G1 stars at [Fe/H]$\sim-1.6$ is in good agreement with \cite{myeong19}.The \textit{Gaia}-Sequoia G2 stars appear to have a separate [$\alpha$/Fe] knee at a much lower metallicity, [Fe/H]$\sim -2.3$, but with high uncertainty.

Data from the Sculptor dwarf galaxy (considered a ``textbook dwarf spheroidal galaxy'', \citealt{hill19}) is compared with the SB02 data in Fig.~\ref{fig:sculptmgca}. The [$\alpha$/Fe] knee in the Sculptor data is near [Fe/H]$\sim-1.8$, between our identifications for the \textit{Gaia}-Sausage and \textit{Gaia}-Sequoia G2 groups. 
Interestingly, [Ca/Fe] in the G2 stars aligns well with [Ca/Fe] in Sculptor (see Fig.~\ref{fig:sculptmgca}).  This agreement is not present in the [Mg/Fe] abundances though, which may point to differences in the star formation histories, e.g., similar yields of Mg and Ca from Type II supernova, but additional Ca in late contributions from Type Ia supernovae.  Drawing tentative conclusions from Figs.~\ref{fig:capturedmgca} and \ref{fig:sculptmgca}, the \textit{Gaia}-Sausage and G2 \textit{Gaia}-Sequoia stars occupy independent chemo-dynamical space, yet with star formation histories that resemble those of the low mass dwarf galaxies.

Finally, one of the Mg-poor stars in our ab initio subset, G184-007, stands out from the majority of \textit{Gaia}-Sequoia G2 stars in Fig.~\ref{fig:sculptmgca}. Both G184-007 and the chemically peculiar star G251-024, appear much lower in [Mg/Fe] than similar stars in the Gaia satellite galaxies or Sculptor. Sub-solar [Mg/Fe] values in stars are rare, however similar stars have been found in the nearby dwarf galaxies, e.g., the Carina and Sextans dwarf galaxies \citep{nor17, jablonka2015, venn12}, the Tri II ultra faint dwarf galaxy \citep{venn2017}, as well as the unusual star cluster NGC~2419 \citep{cohen2012}.  In NGC~2419, the low Mg abundances is anti-correlated with other elements (K and Sc), but there is no relationship with Ca.  This unusual chemical pattern, combined with its highly retrograde orbit, convinced \cite{cohen2012} that NGC 2419 is not a globular cluster, but rather the dense core of an accreted dwarf galaxy.   As the Mg-poor (Ca-normal) star G184-007 (G251-024 is discussed further in Section~\ref{sec:g251}) is also on a highly retrograde orbit, this could indicate similarities in the star formation history between the textit{Gaia}-Sequoia G2 and NGC~2419 progenitors.

\subsubsection{\label{sec:chemsig2}Other Chemical signatures of accretion}

\begin{figure}
    \centering
    \includegraphics[scale=0.4]{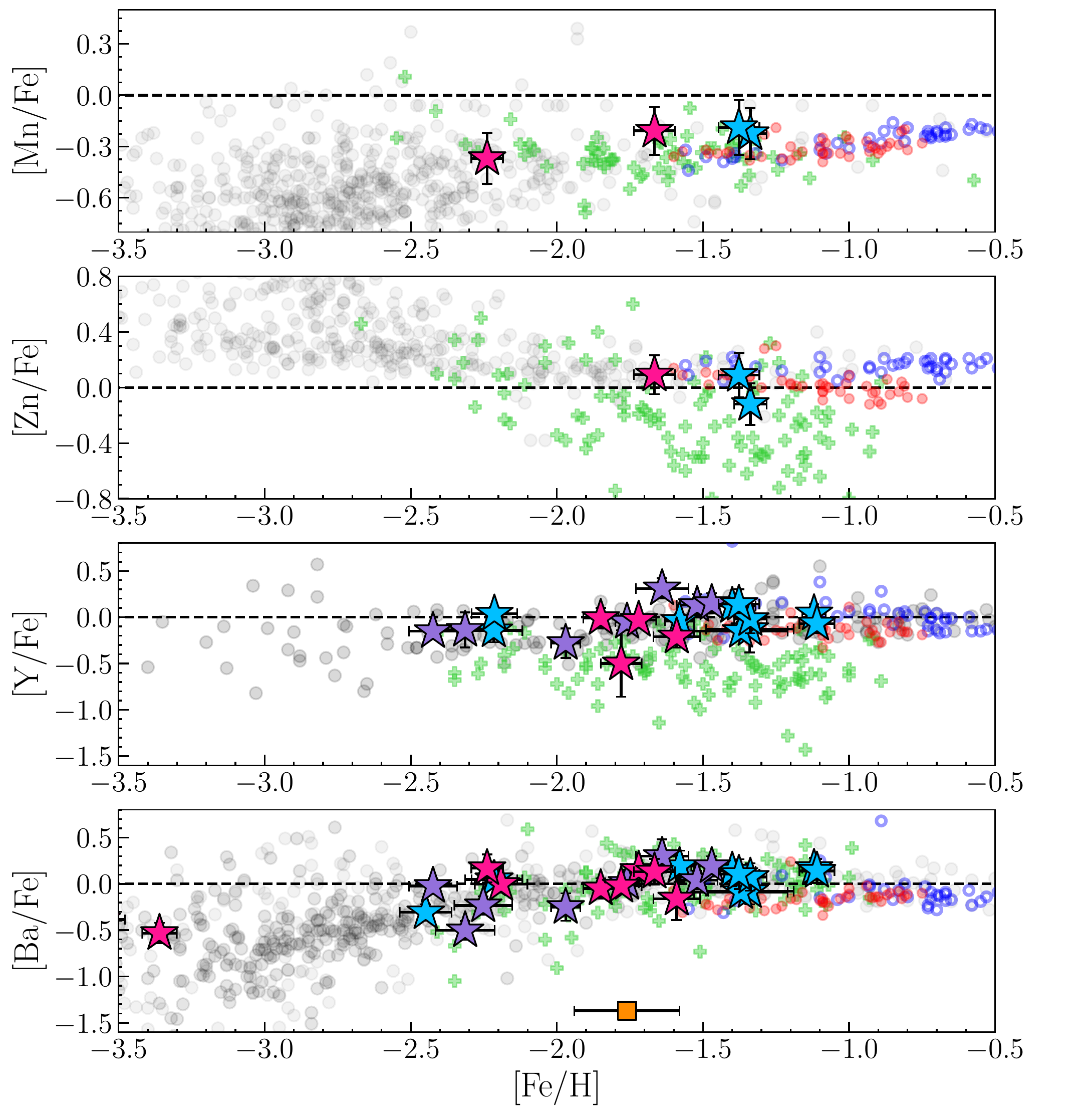}
    \caption{The chemical abundances for seleted elements; Mn, Zn, Y and Ba.  Only the stars in our ab initio re-analysed subset have measured values of Mn and Zn, as these elements were not examined in the original SB02 study. Literature data (black dots, red and blue circles) follows the convention in Fig.~\ref{fig:sb02ns1011_mgca}. Possible members of the \textit{Gaia}-Sequoia accretion event are shown as pink (high orbital energy, G2) and purple (low orbital energy, G1) stars, while stars associated with the \textit{Gaia}-Sausage accretion event are shown in sky blue. The chemically distinct star G251-024 (discussed in Section \ref{sec:g251}) is shown as the orange box in [Ba/Fe]. Data for the Sculptor dwarf galaxy is also shown for Mn from \citep{north2012}, Zn from \citet{skul17, hill19}, Y from \citet{skul2019} and Ba from \citet{hill19}.
    \label{fig:mnznba}}
\end{figure}

In Fig.~\ref{fig:mnznba}, we show the abundances for a select set of elements; the neutron-capture elements Ba and Y, the odd-Z element Mn, and the low condensation temperature element Zn. 
These four elements have been linked to the accretion of stars from dwarf galaxies with distinct star formation histories.  For example, low Mn has been associated with stars accreted from the Sagittarius dwarf galaxy and other satellite remnants \citep[e.g.,][]{mcw2003, north2012, hasselquist2017}. Low Mn may reflect a lower number of high mass star polluters (in elements that form during explosive nucleosynthesis via Type II supernovae), although, it has also been proposed that the double vs single degenerate SN Ia yields may affect Mn abundances in dwarf galaxies \citep{delos2020}. Similarly, Zn and the light neutron capture element, Y have been found to be systematically under abundant in many intermediate-metallicity stars in the classical dwarf spheroidal galaxies \citep{berg15, skul17, hill19, skul2020}. The heavy neutron-capture elements also have a complex chemical evolution history in dwarf galaxies due to metallicity dependent yields from massive stars and AGB, as well as stochastic sampling of the IMF and inhomogeneous mixing in many dwarf galaxies \citep[e.g.,][]{venn04, tolstoy09, letarte10, skul2020}.   Nevertheless, our abundances of these elements in the stars we associate with  \textit{Gaia}-Sausage and \textit{Gaia}-Sequoia do not show any significant offsets relative to the normal Galactic halo stars.

The Y abundances of the stars that may be associated with the Gaia dwarf remnants do not display the low Y abundances seen in the Sculptor stars, following the halo star Y-distribtion more closely. However, the Ba abundances of the potential dwarf remnant stars do show agreement with the Ba abundances of both MW halo or Sculptor stars. \cite{skul2020} suggest that the disagreement between the Y and Ba abundance trends seen in Sculptor at metallicites greater than [Fe/H]$\sim-2$, is evidence of two distinct production sights. We don't see clear evidence for this in our potentially captured objects. There are no r-process rich stars in this sample, nor stars that are clearly low in neutron-capture elements.   The latter are common in ultra faint dwarf galaxies \citep[e.g., ][]{ji19}, whereas r-II stars have only been seen in a few dwarf galaxies (Ret II and Tuc III; \citealt{ji16} and \citealt{hansen2017}).

\subsection{Other interesting stars}
In this section, we discuss other interesting stars that are not kinematically associated with the \textit{Gaia}-Sausage or \textit{Gaia}-Sequoia events.

\subsubsection{\label{sec:g251}The chemically peculiar star G251-024}
\begin{figure}
\centering
    \includegraphics[width=\linewidth]{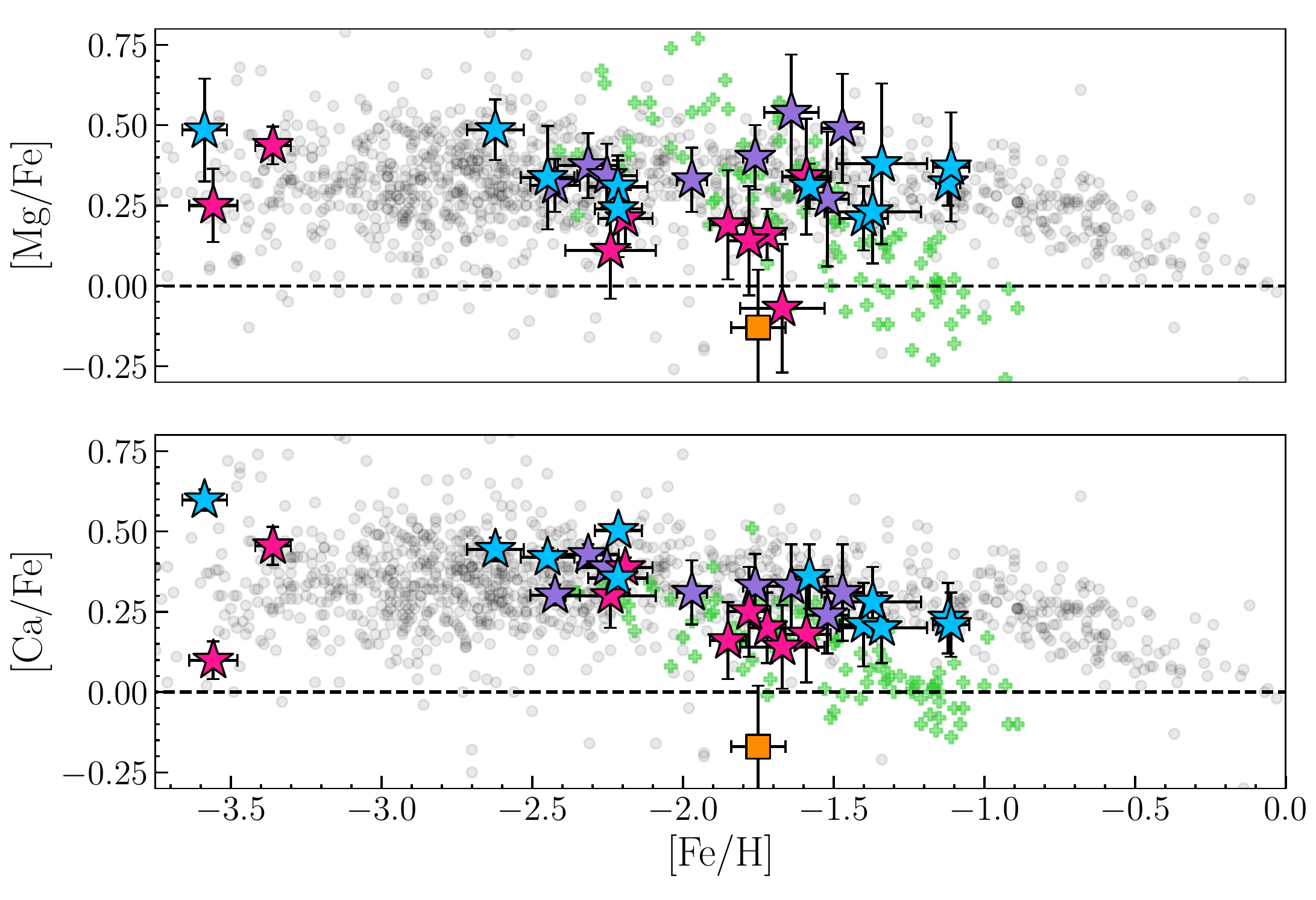}
    \caption{Same as Fig.~\ref{fig:capturedmgca} with the addition of the Sculptor data from \citet{hill19} and the chemically peculiar star G251-024.} 
    \label{fig:sculptmgca}
\end{figure}

G251-024 is also known is BD +80$^{\text{o}}$245, previously discovered and analysed by \cite{fulbright02} and \cite{ivan03}. Both groups found low $\alpha$-element and Ba abundances, which we confirm in this study ([Mg/Fe]=$-0.11\pm0.18$, [Ca/Fe]=$-0.19\pm0.19$, [Ti/Fe]=$-0.27\pm0.11$, and [Ba/Fe]$= -1.37\pm0.18$). \cite{ivan03} also found it is low in r-process elements, e.g., [Eu/Fe]$=-0.64\pm0.18$.  This star is noted in Fig.~\ref{fig:sculptmgca} as an orange square, to show how distinct its chemistry is from the main sample and all other chemo-dynamical groups. This star was also examined in \cite{venn12} where it was compared to the chemically peculiar star Car-612 found by \cite{venn12} in the Carina dwarf galaxy.  It was proposed then that these stars are unusual due to enhancements in the iron-group elements, e.g., possibly forming in a pocket of SN Ia products that dilutes [X/Fe] abundances locally.  This was further explored for a larger number of stars in Carina by \cite{nor17}, and implies inhomogeneous mixing of the interstellar medium at early times in low mass dwarf galaxies.  

Unfortunately this star has the worst astrometry in the entire SB02 sample. The Gaia parameters,  \texttt{astrometric\_chi2\_al} and \texttt{phot\_bp\_rp\_excess\_factor} have values of $\sim2311$ and $\sim1.49$ respectively, restricting any further chemo-dynamic analyses.

\subsubsection{Li in G037-037}
G037-037 is the only star in our sub-sample hot enough (6000~K) to have preserved its initial Li abundance, protected from destruction by convection (see \cite{spite82a} and references within).  Spectrum synthesis of the \ion{Li}{I} 6707 \AA\ line (including hyperfine structure components and NLTE corrections) results in A(Li) =  $2.16\pm0.10$ dex.  The Spite Plateau \citep{spite82b, spite82a} is identified by \cite{melendez04} near A(Li) = 2.37 dex, however in the metallicity range ([Fe/H] $\sim -2.5$) of turn-off stars (log$g$ $\ge 3.7$) like G037-037, the Lithium Plateau is identified by \citet[][and references therein]{aguado2019} at A(Li)=2.2 dex, in good agreement with our result.

\subsubsection{G122-051: A metal weak disk star?}
G122-051, a member of the original SB02 sample, is a prograde star in a highly elliptical, planar orbit (Z$_{\rm max}<0.17$ kpc, R$_{\rm apo}=19.6$ kpc, and $e$ = 0.87). Using the solar values stated in Section \ref{sec:orb}, we find that the azimuthal velocity of G122-051 is +108 kms$^{-1}$. Comparing this azimuthal velocity with that of the three proposed MW components; halo, thick and thin disks and considering its chemistry, [Fe/H]$=-1.34\pm0.06$ and [$\alpha$/Fe] = $0.26\pm0.05$, then G122-051 is not clearly an outer halo star. A better description is that it is a member of the proposed \textit{metal weak thick disc} (MWTD) \citep{chiba2000, beers2002}, where the mean MWTD azimuthal velocity in \cite{carollo10} is $V_{\phi}=100-150$kms$^{-1}$ and mean metallicity [Fe/H]$\sim-1.3$. 

Similarly, the MWTD parameters from \cite{kordopatis13} again associate this star most closely with the MWTD ($V_{\phi}=123\pm16$kms$^{-1}$ and [Fe/H]$\sim-1.6$). Alternatively, using the convention of \cite{hayes18} this star is classified as a high Mg star, ([Mg/Fe]$\sim+0.3$ dex), where its motion also agrees with the high Mg
population ($V_{phi}\sim120-150$kms$^{-1}$). Although, \cite{hayes18} do not claim that their high-Mg population is related to the MWTD, they do suggest that they could be related.

The existence of a MWTD as a separate chemo-dynamic component of the MW thick disc is still the subject of debate. In some studies, a separate metal weak component of the thick disk is needed to fit the observed rotational properties of low-metallicity stars like G122-051 near the plane \citep{carollo10, kordopatis13}. When treated as a discrete component of the MW disk, the MWTD is generally suggested to have formed in-situ, with contributions from mergers dynamically heating the precurser MWTD to the scale height seen today \citep{hayes18, haywoodge, dimatteo18}. 

More recently, \citet{sestito19, sestito2020} have combined the metallicities and radial velocities from the Pristine \citep{aguado2019} and LAMOST \citep{lamost} surveys with Gaia DR2 parallax and proper motion values to find a large number ($\geq300$) of very metal-poor stars ([Fe/H] $\leq$ -2.5 dex) that currently reside in the disk (|Z| $\leq$ 3 kpc) and are confined to the plane of the disk throughout their orbit.  Moreover, this sample suggests prograde motion, implying that they merged into, formed within, or formed concurrently with the Milky Way disk.  This very metal-poor component, in addition to the MWTD, suggests that the history of the disk was quiet enough to allow these stars to retain their disk-like orbital properties, which is a challenge to theoretical and cosmological models.

Finally, in the chemo-dynamic study of MW halo stars, \cite{dimatteo18}, found that only about 25\% of stars in the metallicity range $-1.5\leq$ [Fe/H] $\leq-1.0$ and Mg-abundance range $0.25\leq$ [Mg/Fe] $\leq0.35$ are accreted stars - suggesting G122-051 formed in-situ. However, they also note the appearance of a peak in the number of stars found near $V_{\phi}\sim100$kms$^{-1}$ (translated to our coordinate system) in the aforementioned metallicity and Mg-abundance range. They suggest this concentration of stars (which again includes G122-051) could either be accreted in an event that results in a variety of deposited orbits, or denote the separation between the MWTD and the remainder of the thick disk.

\section{Conclusions}\label{sec:conc}
We have re-examined the \cite{sb02} (SB02) data set of ``outer halo stars" on energetic or highly retrograde orbits, using Gaia DR2 astrometry for the entire dataset, and high resolution Keck HIRES and Gemini GRACES spectra for an ab initio analysis of a subset of six stars.  Our results are used as test cases and calibrations of the SB02 results. Stellar parameters have been determined using two methods, (1) our \textit{isochrone-mapping} method using Gaia DR2 data and the Dartmouth Stellar Evolution Database, and (2) a classical 1DLTE model atmospheres analysis using spectroscopic indicators. We find excellent agreement for the stellar parameters from both methods and with SB02 when metallicity [Fe/H]$>-2$.  For lower metallicity objects, we suggest generic stellar isochrones may be insufficient for our \textit{isochrone-mapping} method. We carry out a 1D~LTE abundance analysis for all of the metal-poor stars in SB02 ([Fe/H]$\le-2$) using the published EWs. Some adjustments in metallicity [Fe/H] are noted, but generally, our new abundance ratios [X/Fe] are in very good agreement with SB02.  Exceptions from our ab initio analysis include two stars that appear to be more Mg-poor (G184-007, G189-050), one star that is very Na-poor (G158-100), and others that may be slightly higher in Y and Ba. We also find Li in one star (G037-037) that is consistent with the Lithium plateau, and determine Mn and Zn abundances that were not determined by SB02.

When combined with Gaia DR2 data, we find that 11 stars in the SB02 sample are dynamically coincident with the \textit{Gaia}-Sausage satellite merger, including one very low metallicity star (G238-030) ([Fe/H]$\sim-3.6$). We also find 17 stars that are dynamically coincident with the \textit{Gaia}-Sequoia accretion event, including one very metal-poor star near [Fe/H] $=-3.5$ (G082-023).  Both metal-poor stars have low masses and isochrone ages older than 10 Gyr. When examining the \textit{Gaia}-Sequoia stars in $L_{z}$ vs $E$ space, we find two distinct groups, split into high orbital and low orbital energy stars. A knee in [$\alpha$/Fe] is found for both the \textit{Gaia}-Sausage stars and low orbital energy \textit{Gaia}-Sequoia stars (G1) near [Fe/H]$\sim-1.6$, while a knee in the high orbital energy \textit{Gaia}-Sequoia (G2) stars is tentatively identified near [Fe/H]$\sim-2.3$. These are consistent with other analyses of the [$\alpha$/Fe] knees in these systems based on SDSS APOGEE data. If the metal-poor stars in these samples are true members of the \textit{Gaia}-Sausage and \textit{Gaia}-Sequoia remnants, they present opportunities to probe the low metallicity tail and early star formation history of these systems. 

Additionally, we find that several individual stars have interesting chemo-dynamical properties. These include the two Mg-poor stars in our subset, one of which is dynamically associated with the \textit{Gaia}-Sequoia accretion event (G184-007).  We also find one star that could be part of a metal weak thick disk in the MW (G122-051).  The star G251-024 (also known as BD +80$^o$ 245) is particularly interesting, but its Gaia DR2 astrometry is too poor for an orbital analysis.

The dynamical picture of the MW is currently evolving thanks to the spectacular view from Gaia, while the chemo-dynamic picture of the MW is just emerging in the upcoming era of spectroscopic surveys (SDSS-V, WEAVE, PFS, and 4MOST). Undoubtedly, this combination of detailed chemical abundances and orbital dynamics will provide the best evidence for testing our models of the formation and accretion history of the Milky Way and its satellites in the coming decade.

\section*{Acknowledgements}
We thank the referee for their helpful comments that have improved this paper greatly. We also want to thank GyuChul Myeong and Helmer Koppleman for their suggestions and advice, both of which have had a major impact on this paper. Thanks also to Ken Freeman, John Norris, Rosemary Wyse, Mike Irwin, Luca Casagrande, Anke Arentsen, Andr\'{e}-Nicolas Chen\'{e}, Aaron Dotter and Jo Bovy for their invaluable advice, suggestions, isochrones and comments throughout the process. Thanks especially to Mike Irwin for the initial suggestion to examine the full dynamics of the sample. SM acknowledges the support provided for a portion of this research by the Natural Sciences and Engineering Research Council of Canada (NSERC) Undergraduate Student Research Awards (USRA). 
KV acknowledges funding from the National Science and Engineering Research Council Discovery Grants program and the
CREATE training program on New Technologies for Canadian Observatories.
The authors wish to recognize and acknowledge the very significant cultural role and reverence that the summit of Maunakea has always had within the indigenous Hawaiian community.  We are most fortunate to have the opportunity to conduct observations from this mountain. This research made use of Astropy, \url{http://www.astropy.org} a community-developed core Python package for Astronomy \citep{astropy2018} and SciPy \citep{scipy}. This work is based on observations obtained at the Gemini Observatory, which is operated by the Association of Universities for Research in Astronomy, Inc., under a cooperative agreement with the NSF on behalf of the Gemini partnership: the National Science Foundation (United States), National Research Council (Canada), CONICYT (Chile), Ministerio de Ciencia, Tecnolog\'{i}a e Innovaci\'{o}n Productiva (Argentina), Minist\'{e}rio da Ci\^{e}ncia, Tecnologia e Inova\c{c}\~{a}o (Brazil), and Korea Astronomy and Space Science Institute (Republic of Korea).This research has made use of the NASA/ IPAC Infrared Science Archive, which is operated by the Jet Propulsion Laboratory, California Institute of Technology, under contract with the National Aeronautics and Space Administration. This research has made use of the Keck Observatory Archive (KOA), which is operated by the W. M. Keck Observatory and the NASA Exoplanet Science Institute (NExScI), under contract with the National Aeronautics and Space Administration.  

\section*{Data Availability}
The data underlying this article are available in the article and online through provided links and supplementary material.

%%%%%%%%%%%%%%%%%%%%%%%%%%%%%%%%%%%%%%%%%%%%%%%%%%


 % if your bibtex file is called example.bib

\bibliographystyle{mnras}
\bibliography{Bibliography.bib}

%\begin{thebibliography}{99}
%\end{thebibliography}

%%%%%%%%%%%%%%%%%%%%%%%%%%%%%%%%%%%%%%%%%%%%%%%%%%

%%%%%%%%%%%%%%%%% APPENDICES %%%%%%%%%%%%%%%%%%%%%
%\onecolumn
%\newpage
\appendix

\section{Spectral Analysis Methodology}
\label{app:spectromethod}
\subsection{Colour temperatures and physical gravities}\label{sec:isochronemap}

\begin{table*}
\caption{Photometric and distance information for each star. This includes the Gaia DR2 $G$-band photometry and $BP-RP$ colours, and the 2MASS $J - K_{s}$ colours \citep{2MASSC}. Geometric distances are determined from inverting the Gaia DR2 parallaxes, and Bayesian corrected distances are from  \citet{bailer18}.  The reddening E(B-V) are from the Bayestar19 reddening map \citep{greenred} assuming the geometric distance and the conversion $\text{E(B-V)} = 0.981\times(\text{Bayestar19}$) from \citet{schlafly11}.
\label{tab:photvals}}

\begin{tabular}{@{}lllllll@{}}
\hline
Star     & $G$ 		& $BP-RP$	& $J-K_{s}$			& GDR2 Dist.		 & Geo. Dist. 				& E(B-V)	  \\
		 & mag		& mag 		& mag 				& (pc) 				 & (pc)					   	& 				\\\hline\smallskip
G037-037 & 12.13	& 0.76		& $0.381\pm0.03$	& $295.34\pm6.05$	 & $293.01^{+6.13}_{-5.88}$	& $0.11^{+0.02}_{-0.03}$				\\\smallskip
G158-100 & 14.69   & 0.95		& $0.484\pm0.04$ 	& $461.43\pm10.5$    & $455.53^{+4.62}_{-15.97}$ & $0.01^{+0.02}_{-0.01}$ 			\\\smallskip
G184-007 & 14.19   & 1.09		& $0.453\pm0.04$	& $315.37\pm1.87$  	 & $312.53^{+1.86}_{-1.84}$ & $0.10^{+0.01}_{-0.03}$				\\\smallskip
G189-050 & 12.49   & 0.95		& $0.462\pm0.02$ 	& $194.02\pm1.54$  	 & $192.95^{+1.55}_{-1.52}$ & $0.00^{+0.02}_{-0.00}$	\\\smallskip
G233-026 & 11.68	& 0.91		& $0.466\pm0.03$ 	& $137.84\pm0.46$	 & $137.30\pm0.46$ 			& $0.00^{+0.01}_{-0.00}$	\\\smallskip			
G262-021 & 13.58	& 1.04		& $0.521\pm0.05$ 	& $271.78\pm0.96$	 & $269.66\pm0.95$			& $0.04^{+0.11}_{-0.02}$		\\\hline
\end{tabular}
\end{table*}

Physical gravities and colour temperatures are determined simultaneously, using a Monte Carlo (MC) exploration of the Gaia DR2 stellar magnitudes, parallaxes and reddening coupled with \texttt{DSED} \citep{dsed} isochrones.  This method is entirely independent of spectroscopic methods, other than the initial assumption of the metallicity and [$alpha$/Fe] to create the stellar isochrones.  An age and mass is also assumed for our stars to build the isochrones and break any potential dwarf-giant degeneracy (we assume 12 Gyr \citep{car16} and 0.8 M$_{\rm sun}$.) Isochrones were constructed using both the Gaia DR2 and 2MASS \citep{2MASSC} filters to create $G$ vs $BP-RP$ and $J$ vs $J-K_{s}$ CMDs from which to map stellar parameters from the isochrones onto our stars.

\begin{figure}
\centering
\includegraphics[width=\linewidth]{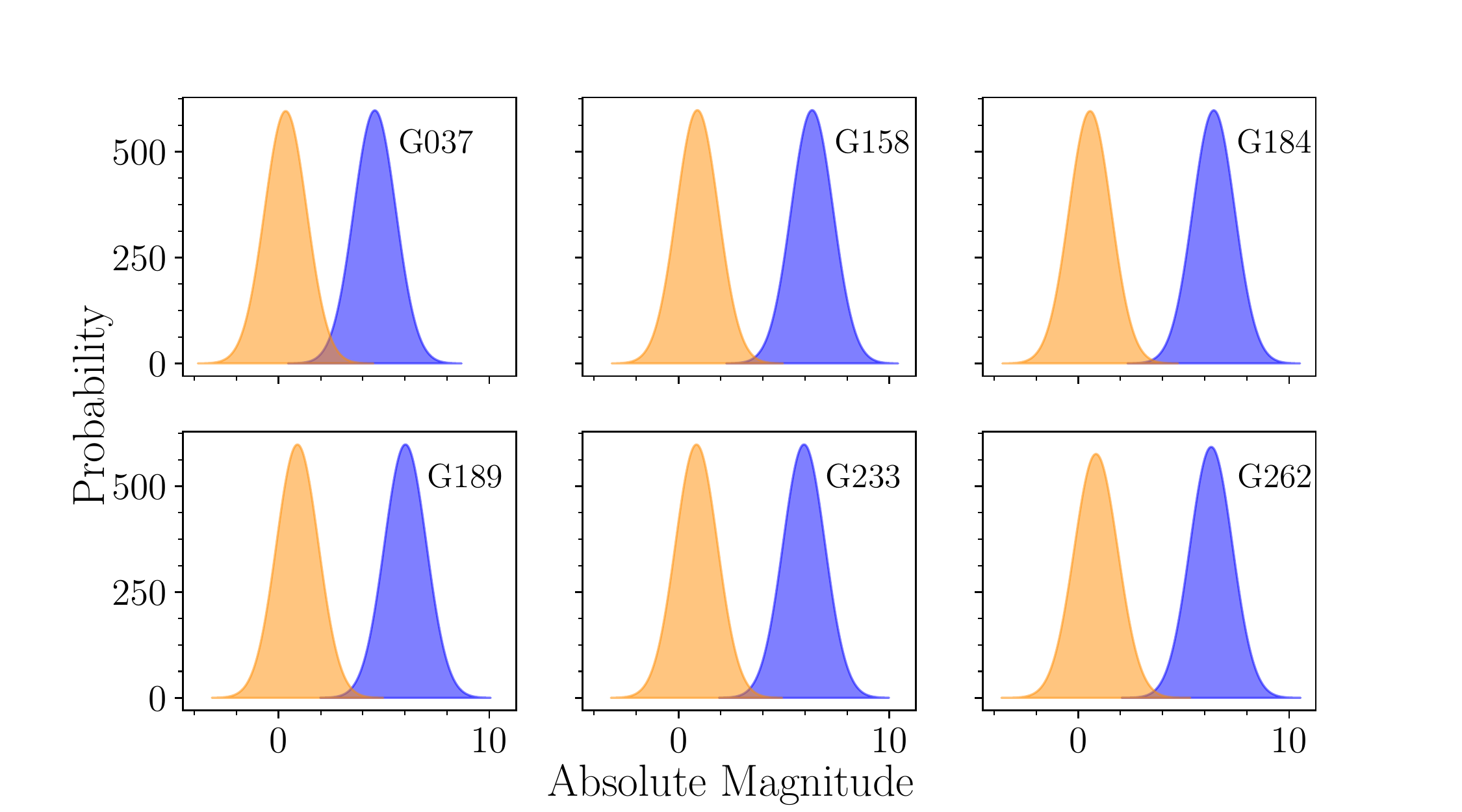}
\caption{Probability density distributions of absolute $G$ and $BP-RP$ magnitudes for each star following 1500 realizations exploring photometric, reddening and parallax errors. The blue distribution shows the spread in the absolute Gaia $G$ band magnitude, while the orange distribution shows the spread in absolute Gaia $BP-RP$ magnitude.}
\label{fig:pdfs} 
\end{figure}

For the MC estimates, we randomly sample the apparent $G$, $BP-RP$, $J$ and $J-K_{s}$ magnitudes, value of $E(B-V)$, and parallax ($\bar{\omega}$) within their symmetric error distributions.  Reddening corrections are determined using the geometric distances based on Gaia DR2 parallaxes \citet{bailer18}, and the Bayestar19 reddening map from \citet{greenred}.  Corrections for $E(B-V)$  were applied to the Gaia and 2MASS filters using the coefficients from \citet{greenred} and \citet{casa18}, shown in Table \ref{redcoeff}. In the case of non-symmetric error distributions, we assume the larger error value to create a symmetric distribution.   After 1500 realizations, the peak and spread in the probability density distributions were used to place each star on the $G$ vs $BP-RP$ and the $K_{s}$ vs $J-K_{s}$ CMDs, as shown in Fig.~\ref{fig:pdfs}. Simultaneous fitting is performed in both colours to mitigate the effects of uncertainties in the reddening corrections.

\begin{table}
\caption{\label{redcoeff} Reddening coefficients used during the investigation of log $g$ as described in Section \ref{sec:isochronemap} and applied as follows: $m_{\xi,0}=m_{\xi}-R_{\xi}E(B-V)$ where the value $E(B-V)$ is discussed in Section \ref{sec:isochronemap} and the reddening coefficient $R_{\xi}$ for each filter $\xi$ are the tabulated values.}
\begin{tabular}{@{}lll@{}}
\hline
Bands     & Coefficients $R_{\xi}$ & Source \\ \hline
($J, K_{s}$)  & (0.793, 0.303) 	   & \citep{greenred} \\
($G, G_{BP}, G_{RP}$) & (2.740, 3.374, 2.035)	& \citep{casa18} \\ \hline
\end{tabular}
\end{table}

\begin{figure}
\centering
\includegraphics[width=\linewidth]{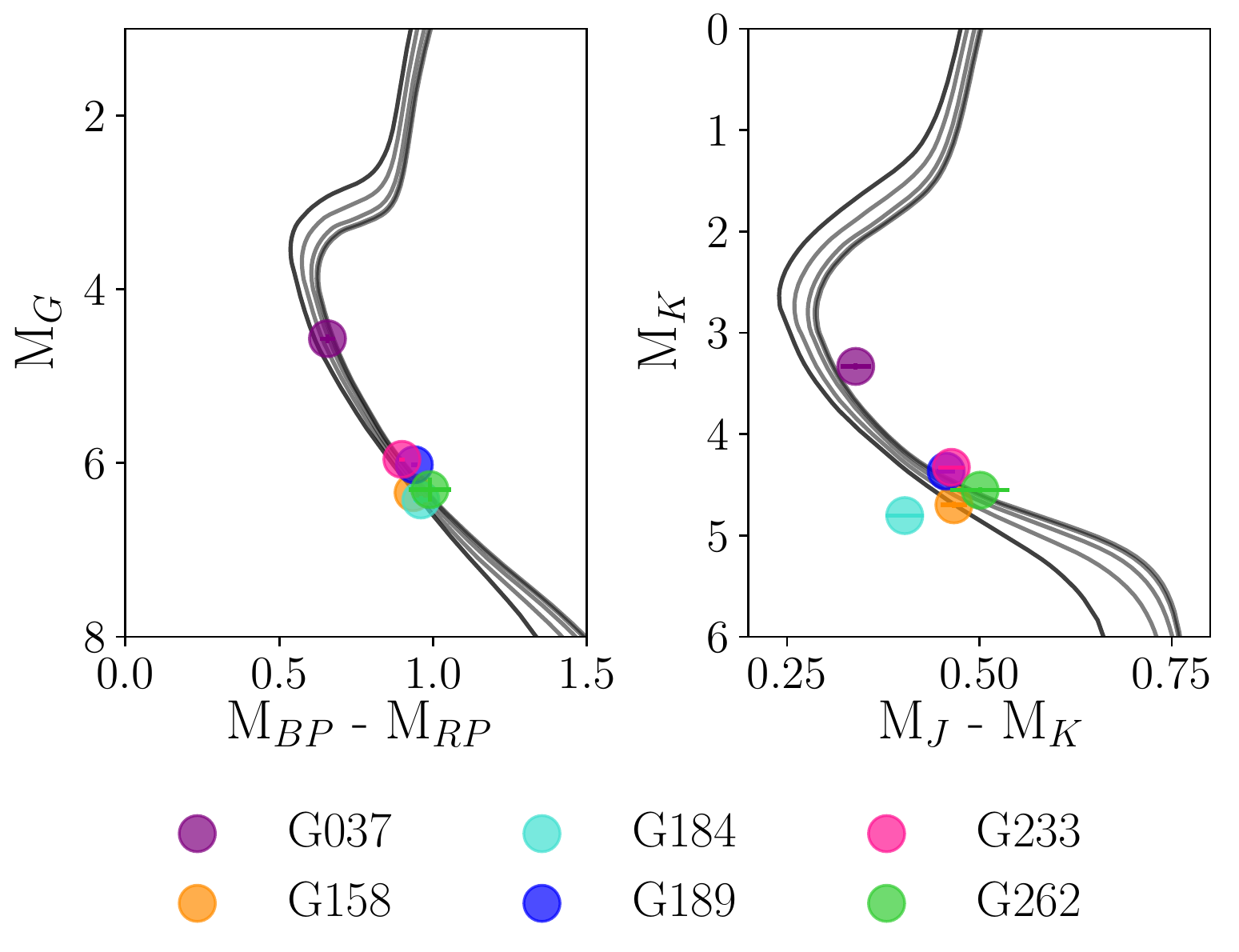}
\caption{Optical and near-IR CMDs of the six stars under study following an MC analysis of their Gaia and 2MASS magnitudes, distances and reddening values. The underlying \texttt{DSED} isochrones span a range of metallicities ($-2.3<$[Fe/H]$<-1.3$), for a fixed alpha enrichment of +0.2 dex and age of 12 Gyr.} 
\label{fig:distcmds} 
\end{figure}

The closest points on the isochrones were then mapped for each star in $G$, $BP-RP$, $J$ and $J-K_{s}$ to determine the physical gravities and colour temperatures. These results are shown in Fig.~\ref{fig:distcmds}, where all of the best-fit isochrones are plotted alongside the stars. Note that a small shift of +0.03 magnitudes was applied to the $BP-RP$ colours of the isochrones to better fit the data prior to mapping. To validate these stellar parameters and account for uncertainties in the underlying isochrone physics, we also varied the isochrone ages by $\pm 2$ Gyr, alpha abundances by $\pm0.2$ dex, and iron abundance by $\pm 0.15$ dex (which is $\sim$1$\sigma$(FeI$_{\rm NLTE}$)).  This analysis proved that the largest uncertainties in the physical gravities and colour temperatures of these stars was due to our assumption of age. We adopt the uncertainty associated with the assumption of age as the uncertainties in our physical stellar parameters; in fact, these uncertainties are very similar to the sum of all of the errors when added in quadrature because the remaining uncertainties are small.

\subsection{\label{sec:spectromethod} Spectroscopic analysis}

\subsubsection{\label{sec:ewmeas} Line Lists and Equivalent Widths}
All spectra were radial velocity corrected using the \texttt{IRAF} task \texttt{fxcorr} using a template synthetic spectrum with similar atmospheric parameters.  All spectra were also continuum normalized using a k-sigma clipping algorithm \cite[e.g.][]{venn12}, therefore the continuum fitting and radial velocity corrections available in DAOSpec were not enabled. 
 
\begin{figure}
\centering
\includegraphics[scale=0.6]{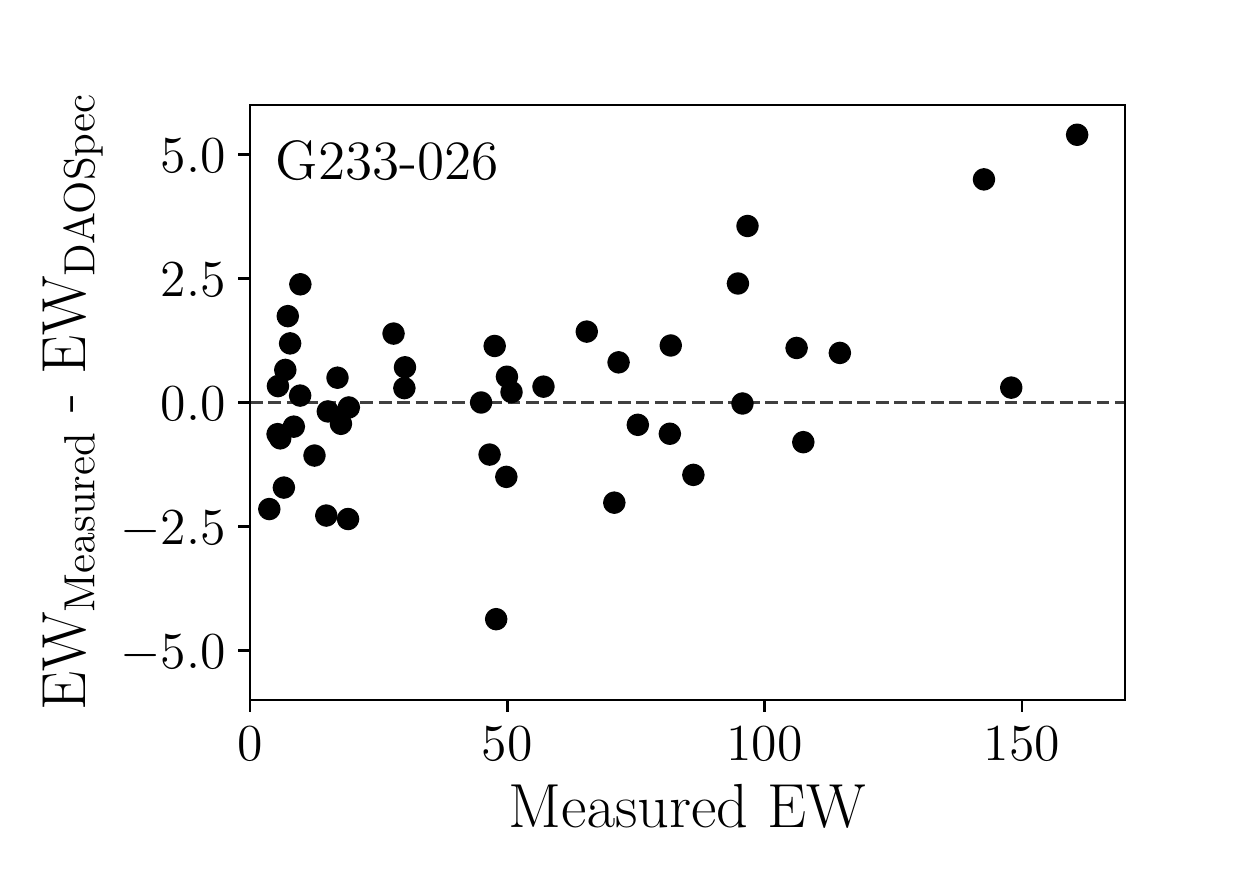}
\caption{Results from an exploration of the fidelity of \texttt{DAOSpec} EW line measurements as described in Section \ref{sec:ewmeas} for the star G233-026. }
\label{fig:ew_comp}
\end{figure}

Initial equivalent width (EW) measurements were made using \texttt{DAOSpec} \citep{daospec}, which finds and fits a Gaussian function to each line in a spectrum for a given line list. EW measurements were also made by hand for $\sim100$ spectral lines ranging from 5-160 m\AA\ using the \texttt{IRAF} task \texttt{splot}.  The results of this comparison are shown in Figure \ref{fig:ew_comp}. Good fidelity was demonstrated in the EW regime from 10 to 150 m\AA.  Deviations in the stronger lines are due to their non-Gaussian profiles (Lorentz wings), but we choose not to include strong lines which are more dependent on precision microturbulence values.  In some cases, individual lines with EW $\le$ 10 m\AA\ were examined and added to the line list dependent on the local SNR in the wavelength region. Line measurements were taken when the local SNR $\ge$ 30.  The lowest EWs are taken as 5 m\AA\ in the best SNR regions ($\ge$ 100).

\subsubsection{\label{sec:specparams} Spectroscopic stellar parameters}
Although we chose to adopt the stellar parameters associated with \textit{isochrone-mapping} method described in Section \ref{sec:isochronemap} for the six stars in our subset, we also derived the stellar parameters spectroscopically to compare the two techniques. This was done through an iterative optimization technique using the LTE line analysis code MOOG \citep{moog}. A temperature range of $\pm$500~K and log\,$g$ of $\pm$0.5 (units of cm/$s^{2}$) were examined around the initial SB02 stellar parameters, in units of $\pm$100~K and $\pm$0.1, respectively. Minimization of log(\ion{Fe}{I}) vs. excitation potential was used to derive the spectroscopic effective temperature (T$_{\rm eff}$), and minimization of log(\ion{Fe}{I}) vs. log(EW/wavelength) was used to determine a microturbulence ($\xi$) value. Ionization equilibrium between the log(\ion{Fe}{I}) and log(\ion{Fe}{II}) abundances was used to constrain a spectroscopic gravity.  Several iterations of this minimization process were undertaken to avoid a local minimum in parameter space. The final spectroscopic parameters are shown in Table \ref{tab:stellparam}.

\subsubsection{\label{sec:NLTE} NLTE corrections}
\begin{figure}
\centering
\includegraphics[scale=0.47]{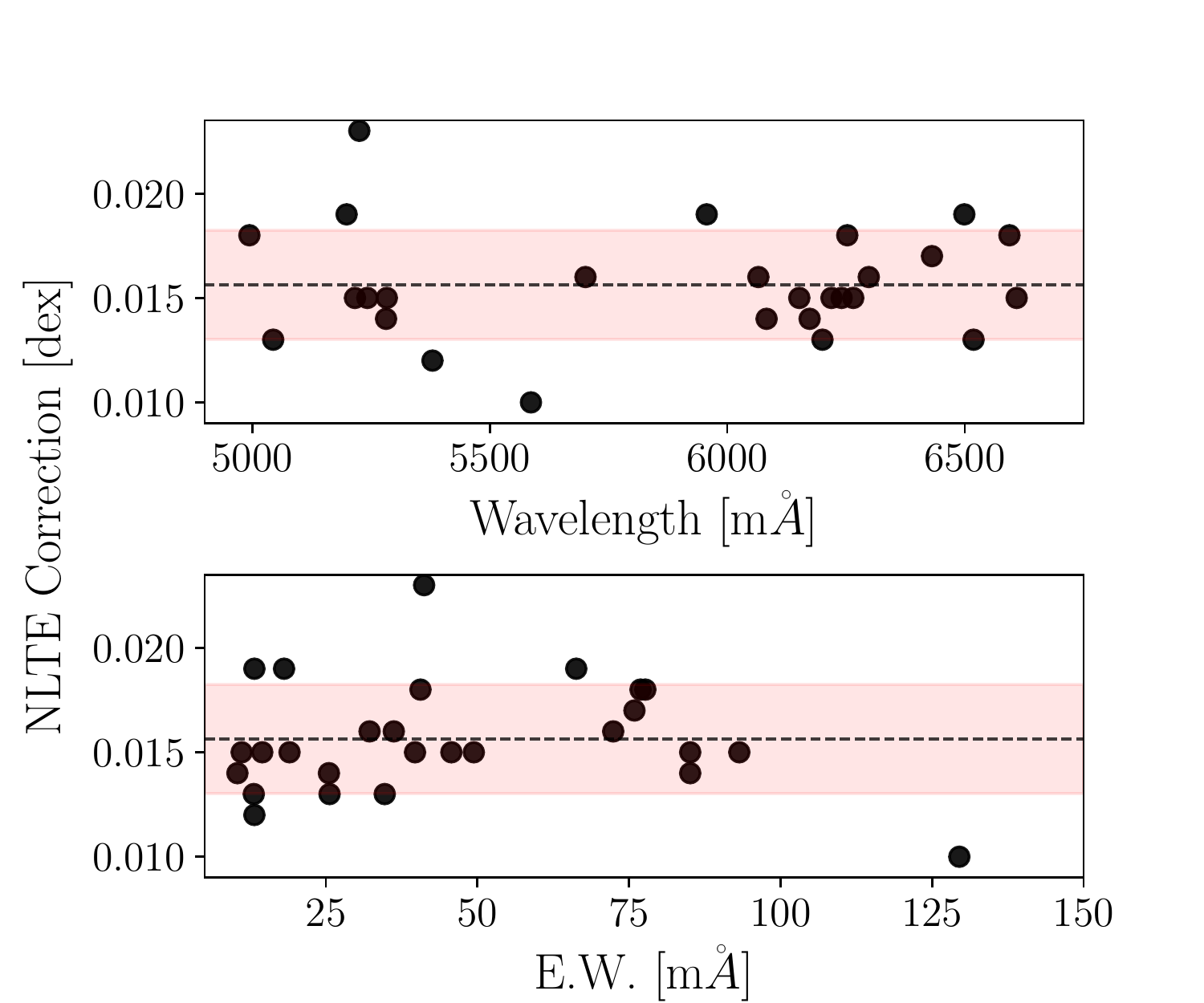}
\caption{\label{nltecor} NLTE corrections for \ion{Fe}{I} lines as a function of wavelength and E.W. measurement for G233-026. The mean NLTE \ion{Fe}{I} correction is shown as a dashed line, with the $1\sigma$ uncertainties in red.}
\label{fig:nlte_cor}
\end{figure}

After adopting the isochrone-mapped stellar parameters we re-derived the metallicity using the 1D, LTE stellar analysis code MOOG \citep{moog}. Following this, we investigated the impact of NLTE corrections for \ion{Fe}{I} lines using the individual spectral line corrections listed in INSPECT \footnote{Data obtained from the INSPECT database, version 1.0 (\url{www.inspect-stars.net})} \citep{bergemann12, lind2012}.  We found that the NLTE corrections for all available lines in our analysis, regardless of EW or $\chi$ value, are similar per star, thus we calculated a simple mean offset to the \ion{Fe}{I} LTE abundances.  Furthermore, the NLTE corrections are generally small (e.g., see Fig.~\ref{fig:nlte_cor}), and have very little impact on our results. The NLTE-corrected metallicities are listed in Table \ref{tab:stellparam} as bolded values. 

For the \ion{Na}{D} lines, the NLTE correction are quite larger, up to $-0.4$ dex, in the lowest metallicity stars in our study. NLTE corrections have been applied to all of our \ion{Na}{I} abundances, including those listed in Table~\ref{tab:abund}.

\subsubsection{Re-examining the Lowest Metallicity Stars in the SB02 Sample}
\label{sec:lowmetbins}
As discussed in Section \ref{sec:sb02gaia}, large disagreements were found between the stellar parameters derived using the \textit{isochrone-mapping} method (Section \ref{sec:isochronemap}) and the original SB02 spectroscopic stellar parameters for stars in the lowest metallicity bins ([Fe/H]$\leq-2$). The disagreement is metallicity dependent, increasing as the metallicity decreases. To investigate this disagreement further, we re-examined the stars in bins (iii) and (iv) to examine the contribution from isochrone choice and the possibility of uncertainties in reddening. A new set of isochrones were created with finer age, metallicity and alpha abundance resolution, exploring a range of ages from 11 to 14 Gyr and metallicities from $-3<[Fe/H]\leq-1.5$. Alternative values of reddening were found using the distance-independent, 2D reddening map of \cite{schlegal98}. The \cite{schlegal98} map was chosen to maintain consistent units with the \cite{greenred} map. Finally, in addition to using the mapping technique we also interpolated the isochrones using a 2D spline interpolation implemented in \texttt{Scipy} \citep{scipy} to compare the two techniques.

The results of the exploration are as follows; (i) low metallicity isochrones ([Fe/H] $\leq$ -2) do not fit the data well for any isochrone age, with the worst disagreement seen on the sub-giant branch, (ii) exploring a larger uncertainty in reddening can lead to a better fit, but is not likely to be the primary cause of disagreement following the first point, and (iii) a better fit was found when the metallicity of the isochrones were increased by 1 dex, on average. Increasing the metallicity to better fit the isochrones also led to better agreement between the spectroscopic and isochrone stellar parameters. Lastly, interpolating the isochrones led to even closer agreement between the two techniques on average.

Despite this investigation into additional sources of error, isochrone parameters in the lowest metallicity bins remain in large disagreement with the spectroscopic parameters. This is a reflection of the isochrone models themselves. In their 2015 paper examining benchmark stars for \textit{Gaia} stellar parameter calibration, \cite{creevey2015} found that existing stellar evolution models could not reproduce the radius, nor effective temperature, of the metal-poor star HD140283 without adjusting the input physics. This was explored further in \cite{joyce18}, where they showed that the implementation of Mixing Length Theory in stellar evolution models, specifically the use of a solar-calibrated mixing length parameter $\alpha_{\mathrm{MLT}}$, does not reproduce fundamental observables of metal-poor stars. Both \cite{creevey2015} and \cite{joyce18} showed that $\alpha_{\mathrm{MLT}}$ must be adjusted to sub-solar values to reproduce observations. Furthermore, \cite{joyce18} conclude that an \textit{adaptive} mixing length must be implemented in stellar evolution models in the future to better model non-Sun-like stars. From this we can conclude that one should exercise caution when applying a single isochrone set to a diverse sample of stars.

\begin{figure}
\centering
\includegraphics[width=\linewidth]{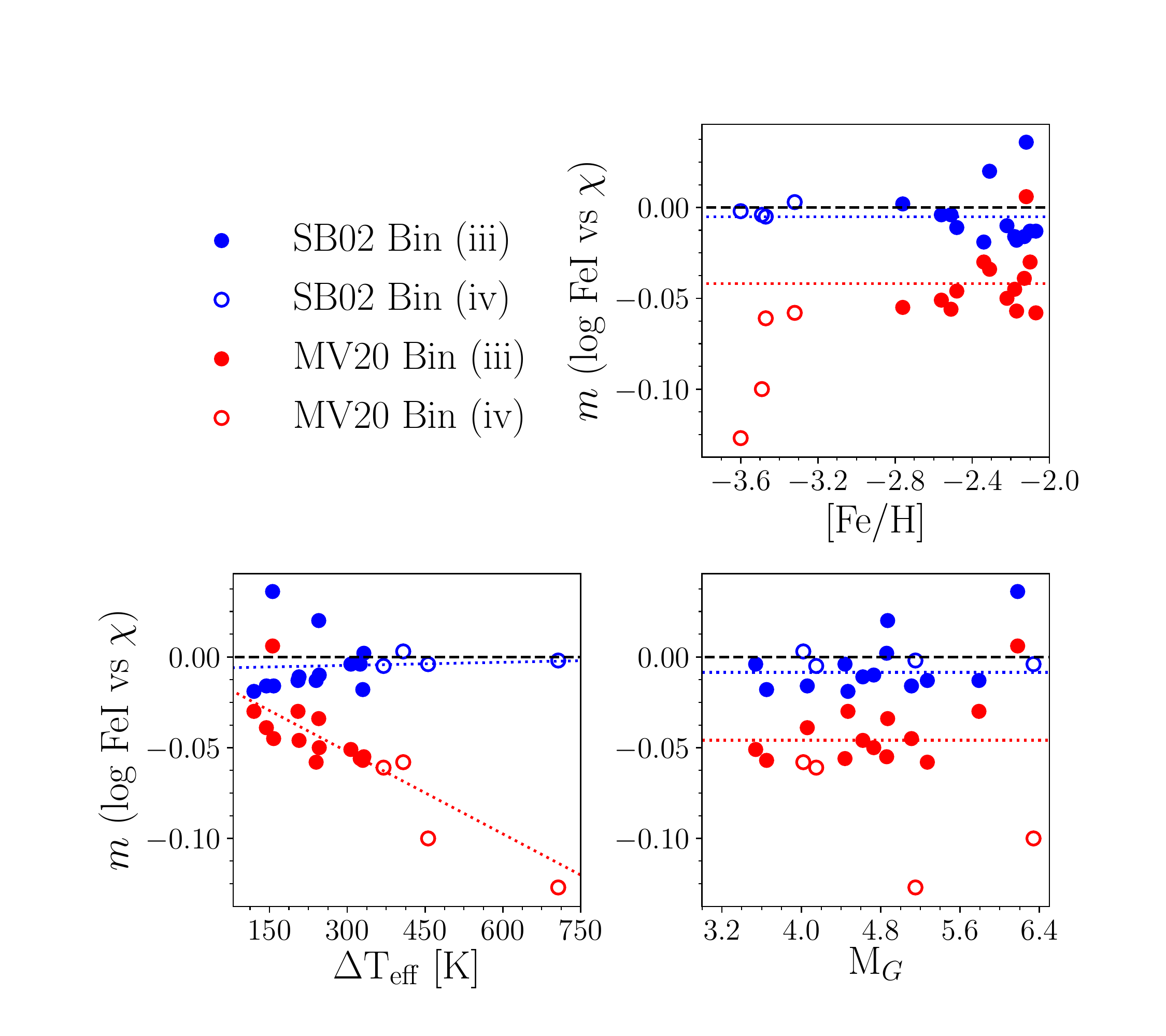}
\caption{Slopes from the linear fit of log(\ion{Fe}{I}) vs. excitation potential as determined by MOOG are shown as a function of: i) (\textit{bottom left}) the disagreement (MV20-SB02) in effective temperature, ii) (\textit{bottom right}) position on the Main Sequence and iii) (\textit{top right} SB02 metallicity. Blue circles represent the slopes found using the original SB02 stellar parameters, red circles represent the slopes found using the isochrone stellar parameters. Filled circles denote bin (iii) stars wile open circles denote bin (iv) stars.}
\label{fig:moogslopes} 
\end{figure}

\begin{figure*}
\centering
\includegraphics[width=\linewidth]{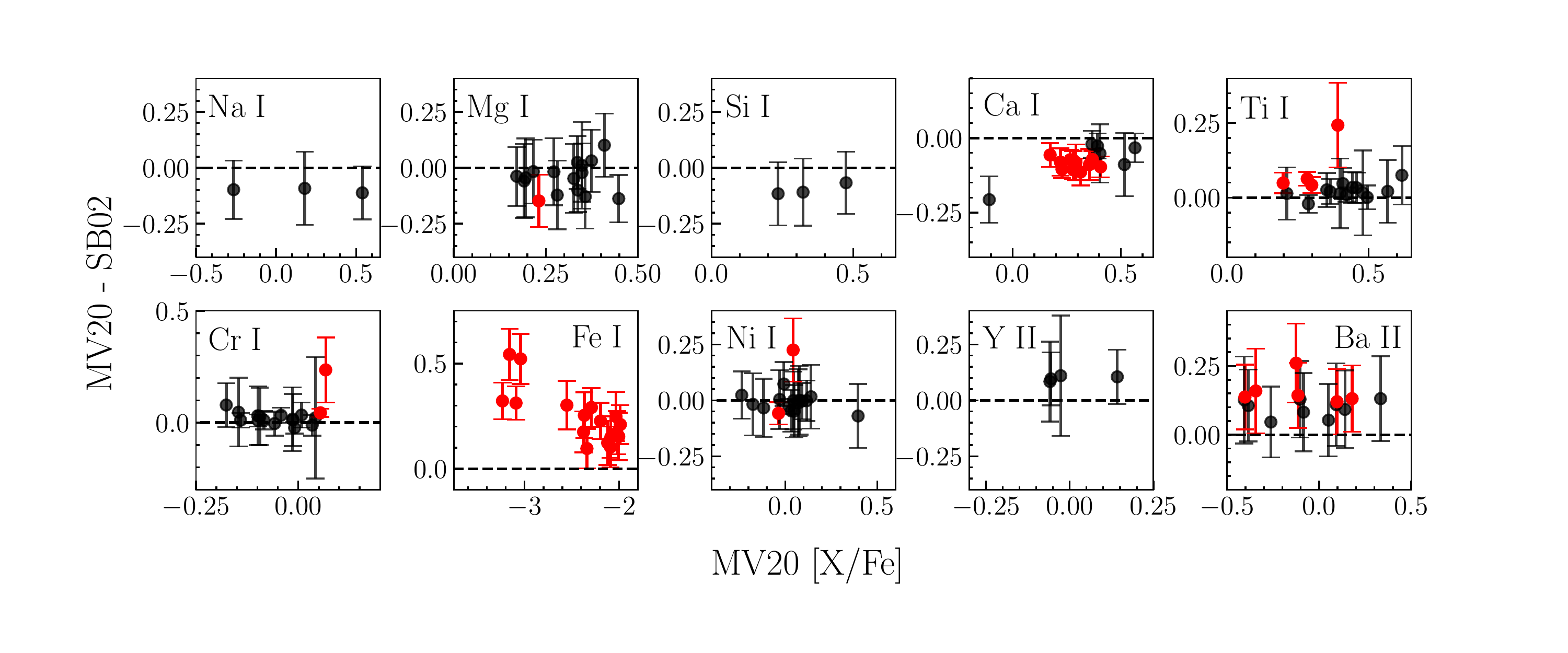}
\caption{Differences in abundances (MV20- SB02) determined for the elements in common between studies. The x-axis is [X/Fe], except for \ion{Fe}{I} where [\ion{Fe}{I}/H] is used. Abundances were determined using the best-fit isochrone stellar parameters, SB02 abundances were re-determined using the original SB02 stellar parameters and our updated linelist. Red points denote abundance differences greater than $1\sigma$.}
\label{fig:updatedabund} 
\end{figure*}

Characteristics of the best-fit isochrones and updated offsets between the two studies are shown in Table~\ref{tab:updatedisooffset}. To examine the effects of the remaining disagreements on the stellar abundances, we re-determined the abundances for stars in bins (iii) and (iv) using both the original SB02 stellar parameters, and the best-fit isochrone parameters. This was done by cross-matching the lines in common between this study and that of SB02, adopting the original EWs of SB02, and updating the atomic data using our modern linelist. We followed the same methodology for the creation of stellar atmospheres and relative abundance determinations, described in Section~\ref{sec:stellabund}. The results of this are summarized in Table~\ref{tab:bins34stellarparams} and Fig.s~\ref{fig:moogslopes} and \ref{fig:updatedabund}.

\begin{table}
\caption{Best-fit isochrone details and updated average differences in effective temperature and surface gravity between this study (``MV20'') and that of SB02 for the lowest metallicity bins.}
\label{tab:updatedisooffset}
\centering
\begin{tabular}{@{}llllll@{}}
\hline
Bin     & Age & [Fe/H]  & [$\alpha$/Fe] & $\Delta$T$_{\text{eff}}$ [K] & $\Delta$log $g$ \\ 
    & Gyr   &   &   & (MV20-SB02)   & (MV20-SB02) \\\hline
iii & 12    & -1.5  & +0.2  & $+232\pm72$ & $0.43\pm0.16$ \\
iv      & 12.5  & -2.5  & +0.2  & $+485\pm131$ & $0.79\pm0.31$ \\
\hline
\end{tabular}
\end{table}

Fig.~\ref{fig:moogslopes} shows the slopes determined by MOOG from linear fits to log(\ion{Fe}{I}) vs. excitation potential $\chi$ for stars in bins (iii) and (iv). Recall that a \textit{good} value of effective temperature determined from 1D~LTE should minimize this slope. It's clear from the left-most plot in Fig.\ref{fig:moogslopes} that as effective temperature disagreements increase the slopes worsen. In other words, the slopes worsen because the isochrone temperatures become hotter. This trend first exceeds the 1$\sigma$ errors in the bin (iii) stars and continues to the lower metallicity stars in bin (iv).
Hence, these differences in the stellar parameters are significant only for the very metal-poor stars, when [Fe/H]$<-2$.  This is important to note as the \textit{isochrone-mapping} method is a convenient way to determine stellar parameters in very metal-poor stars when there are fewer high-quality lines of \ion{Fe}{I} and \ion{Fe}{II} for an accurate stellar parameter determination  \citep[e.g.,][]{Venn20}.

\begin{table}
\caption{Stellar parameters derived using the best-fit isochrones for bins (iii) and (iv) as listed in Table \ref{tab:updatedisooffset}. The difference between isochrone (``MV20'') and original SB02 stellar parameters are listed alongside the parameters.}
\label{tab:bins34stellarparams}
\begin{tabular}{@{}llllll@{}}
\hline
Star		&	Bin	& $T_{\mathrm{eff}}$ [K]	& $\Delta T_{\mathrm{eff}}$ [K]	& log $g$	& $\Delta$log $g$ \\
			& 		&                           & (MV20-SB02)		
&			& (MV20-SB02)	\\\hline
G011-044	& iii	& 6170  & 246   & 4.44		& 0.62 \\
G020-008	& iii	& 6060  & 120   & 3.95		& 0.04 \\
G026-012	& iii	& 6296  & 207   & 4.45		& 0.41 \\
G088-032	& iii	& 6443  & 307   & 4.08		& 0.54 \\
G110-034	& iii	& 5926  & 240   & 4.56		& 0.45 \\
G144-028	& iii	& 5514  & 204   & 4.63		& 0.44 \\
G165-039	& iii	& 6448  & 330   & 4.12		& 0.59 \\
G171-050	& iii	& 6228  & 332   & 4.50		& 0.47 \\
G201-005	& iii	& 6343  & 325   & 4.41		& 0.62 \\
G239-026	& iii	& 6020  & 158   & 4.64		& 0.33 \\
G242-019	& iii	& 5195  & 159   & 4.69		& 0.46 \\
G246-038	& iii	& 5302  & 245   & 4.68		& 0.44 \\
LTT-2415	& iii	& 6439  & 144   & 4.27		& 0.16 \\
G064-012	& iv	& 6444  & 370   & 4.29		& 0.57 \\
G064-037	& iv	& 6530  & 408   & 4.27		& 0.40 \\
G082-023	& iv	& 5390  & 456   & 4.71		& 1.07 \\
G238-030	& iv	& 6090  & 707   & 4.56		& 1.13 \\
\hline
\end{tabular}
\end{table}

Fig~\ref{fig:updatedabund} shows the difference between the stellar abundances determined using our \textit{isochrone-mapping} method and our update for the abundances using the original SB02 stellar parameters. In general, the disagreement in stellar parameters does not result in significant offsets in element abundance ratios, [X/Fe].  Offsets larger than $1\sigma$ are shown in red, and include \ion{Ca}{I}, \ion{Ba}{II}, and some \ion{Ti}{I}. Some of these could simply reflect the smaller error bars associated with elements with a larger number of measured lines.

\begin{table*}
\caption{Sample of the updated abundances for the low metallicity ([Fe/H] < -2) stars in the SB02 sample, bins (iii) and (iv). Abundances were calculated from combining the original SB02 stellar parameters with the updated line list and atomic data used in this study. Uncertainties in abundances were calculated from the line-to-line abundance dispersion ($\sigma_{\text{EW}}$) alone. The full table is included with the online supplementary material.}
\label{tab:bins34abundtab1}
\begin{tabular}{crrrrrrr}
\hline
Star & Bin & [Na/Fe] & [Mg/Fe] & [Si/Fe] & [Ca/Fe] & [Ti I/Fe] & [Ti II/Fe]  \\
\hline
G011-044 & iii & ... & $0.37\pm0.10$ (2) & ... & $0.43\pm0.03$ (17) & $0.34\pm0.03$ (7) & $0.40\pm0.04$ (9) \\
G020-008 & iii & ... & $0.34\pm0.16$ (2) & ... & $0.42\pm0.04$ (17) & $0.41\pm0.04$ (8) & $0.39\pm0.04$ (8) \\
G026-012 & iii & $-0.17\pm0.09$ (1) & $0.37\pm0.09$ (2) & ... & $0.44\pm0.02$ (16) & $0.41\pm0.02$ (6) & $0.46\pm0.03$ (7) \\
G088-032 & iii & ... & $0.49\pm0.09$ (1) & ... & $0.44\pm0.04$ (8) & $0.46\pm0.09$ (3) & $0.50\pm0.09$ (5) \\
G110-034 & iii & ... & $0.29\pm0.10$ (2) & $0.43\pm0.10$ (1) & $0.34\pm0.02$ (18) & $0.15\pm0.03$ (12) & $0.30\pm0.04$ (8) \\
... & ...   & ...   & ... & ... & ... & ... & ...\\
\hline
\end{tabular}
\end{table*}

\begin{table*}
\caption{Sample of the updated abundances for the low metallicity ([Fe/H] < -2) stars in the SB02 sample, bins (iii) and (iv) continued. The full table is included with the online supplementary material.}
\label{tab:bins34abundtab2}
\begin{tabular}{crrrrrrrr}
\hline
Star & Bin & [Cr I/Fe] & [Cr II/Fe] & [Fe I/H] & [Fe II/H] & [Ni I/Fe] & [Y II/Fe] & [Ba II/Fe] \\
\hline
G011-044 & iii & $-0.01\pm0.03$ (6) & $0.22\pm0.10$ (2) & $-2.28\pm0.01$ (58) & $-2.34\pm0.02$ (11) & $0.08\pm0.06$ (7) & $-0.14\pm0.19$ (2) & $-0.50\pm0.10$ (3) \\
G020-008 & iii & $-0.12\pm0.09$ (5) & $0.44\pm0.11$ (2) & $-2.44\pm0.01$ (52) & $-2.47\pm0.03$ (10) & $0.05\pm0.09$ (3) & ... & $-0.31\pm0.09$ (3) \\
G026-012 & iii & $-0.11\pm0.02$ (6) & $0.01\pm0.11$ (2) & $-2.55\pm0.01$ (49) & $-2.59\pm0.03$ (9) & $-0.08\pm0.09$ (4) & ... & $-0.49\pm0.09$ (3) \\
G088-032 & iii & $-0.03\pm0.09$ (4) &... & $-2.62\pm0.02$ (25) & $-2.63\pm0.04$ (9) & $0.12\pm0.09$ (1) & ... &... \\
G110-034 & iii & $-0.03\pm0.04$ (10) & $0.19\pm0.10$ (2) & $-2.16\pm0.01$ (71) & $-2.25\pm0.03$ (11) & $0.08\pm0.10$ (5) & ... & $-0.02\pm0.10$ (3) \\
... & ...   & ...   & ... & ... & ... & ... & ... & ...\\
\hline
\end{tabular}
\end{table*}

\section{More on Orbits}
\label{sec:kinematicdata}

\begin{figure*}
\centering
\includegraphics[scale=0.4]{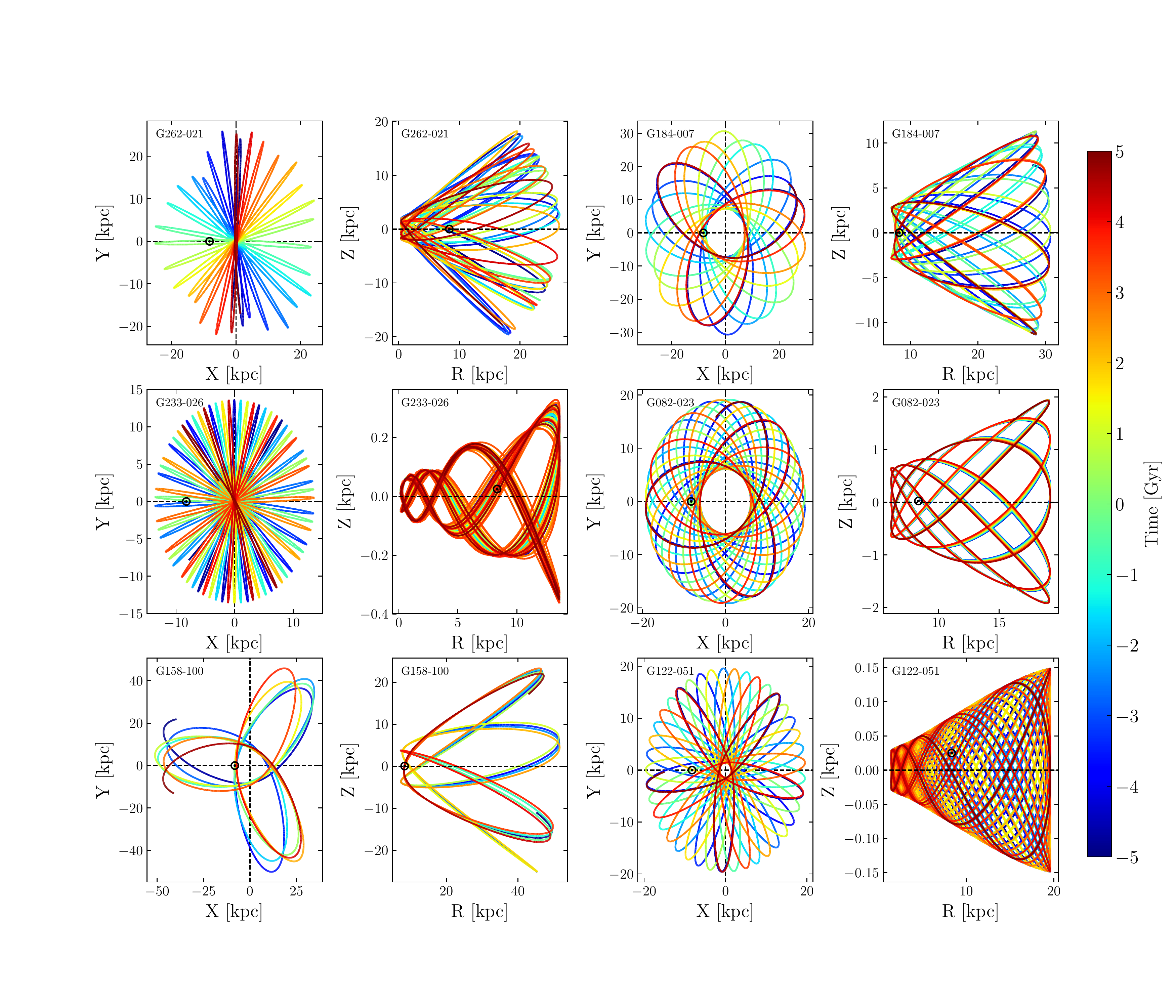}
\caption{Orbits of four stars in our subset (G262-021, G184-007, G233-026 and G158-100) are shown following integration both forwards and backwards in time for 5~Gyr in a \citet{mcmillan17} potential. The first and third columns show galactocentric $X$ and $Y$, and the second and fourth columns show galactocentric cylindrical radius, $R$ and height above the galactic plane, $Z$. The dashed lines mark the zero point of each coordinate, and the black symbol marks the location of the sun.}
\label{fig:star_orbits}
\end{figure*}

As discussed in Section \ref{sec:orb}, orbits were determined for all the stars in the SB02 sample. In Fig.~\ref{fig:star_orbits} we show orbits for four of the stars in our sub-sample (G262-021, G184-007, G233-026 and G158-100) to demonstrate their orbital diversity. An additional two orbits of chemo-dynamically interesting stars are also shown; the metal-poor \textit{Gaia}-Sequoia associated star G082-023 and the potential MWTD star G122-051. The astrometric parameters of the entire SB02 sample are listed in Table~\ref{tab:astroparameters}, resultant orbital parameters are listed in Tables~\ref{tab:orbparameters} and \ref{tab:actionparameters}.  

{\scriptsize
\begin{table*}
\caption{A sample of the basic astrometric parameters from \textit{Gaia} and radial velocities from this work and SB02. The first section shows stars spectroscopically studied in this work with updated radial velocities determined in this work. The remaining radial velocities are from SB02. The full table is included with the online supplementary material.}
\label{tab:astroparameters}
\begin{tabular}{cllllll}
\hline

Star & $\alpha$ & $\delta$ & $\pi$ & $\mu_{\alpha}$ &$\mu_{\delta}$ & RV \\ 
    & [deg] & [deg] & [mas] & [mas/yr]  & [mas/yr]  & [km/s] \\
\hline 
G037-037 & $50.91 \pm 0.06$ & $33.97 \pm 0.03$ & $3.39 \pm 0.07$ & $-72.95 \pm 0.09$ & $-359.86 \pm 0.07$ & $-143.0 \pm 0.4$ \\ 
G158-100 & $8.48 \pm 0.04$ & $-12.13 \pm 0.03$ & $2.17 \pm 0.05$ & $157.6 \pm 0.1$ & $-191.43 \pm 0.09$ & $-360.6 \pm 1.1$ \\ 
G184-007 & $276.05 \pm 0.01$ & $27.29 \pm 0.02$ & $3.17 \pm 0.02$ & $-272.86 \pm 0.02$ & $-169.26 \pm 0.03$ & $-370.6 \pm 0.5$ \\ 
G189-050 & $344.11 \pm 0.03$ & $33.88 \pm 0.03$ & $5.15 \pm 0.04$ & $-98.77 \pm 0.07$ & $-371.24 \pm 0.05$ & $-320.9 \pm 0.6$ \\ 
G233-026 & $339.98 \pm 0.02$ & $61.72 \pm 0.02$ & $7.25 \pm 0.02$ & $-167.54 \pm 0.05$ & $-106.58 \pm 0.04$ & $-313.6 \pm 0.6$ \\ 
G262-021 & $308.86 \pm 0.01$ & $64.9 \pm 0.01$ & $3.68 \pm 0.01$ & $214.19 \pm 0.03$ & $207.98 \pm 0.03$ & $-214.0 \pm 0.5$ \\ 
... & ...   & ...   & ... & ... & ... & ... \\
\hline 

\end{tabular}
\end{table*}}

{\scriptsize
\begin{table*}
\caption{A sample of the orbital properties for the SB02 stars. The first section shows  stars spectroscopically studied in this work. Errors were determined via 100 MC realizations exploring the errors associated with the input Gaia parameters. The full table is included with the online supplementary material.}
\label{tab:orbparameters}
\begin{tabular}{cllllll}
\hline

Star & $R_{\text{peri}}$ & $R_{\text{apo}}$ & $T_{\text{rad}}$ & U  & V & W \\ 
    & [kpc] & [kpc] & [Myr] & [km/s] & [km/s] & [km/s]
\\\hline 
G037-037 & $7.09 \pm 0.08$ & $58.12 \pm 5.67$ & $727.6 \pm 71.5$ & $187.5 \pm 1.11$ & $-298.31 \pm 5.12$ & $-392.11 \pm 8.97$  \\ 
G158-100 & $8.233 \pm 0.001$ & $51.86 \pm 5.7$ & $653.5 \pm 70.1$ & $-42.03 \pm 2.25$ & $-600.88 \pm 13.62$ & $222.29 \pm 3.32$ \\ 
G184-007 & $7.7 \pm 0.01$ & $30.58 \pm 0.75$ & $396.4 \pm 8.4$ & $78.06 \pm 1.69$ & $-569.44 \pm 1.77$ & $165.24 \pm 1.72$ \\ 
G189-050 & $2.1 \pm 0.01$ & $24.09 \pm 0.39$ & $278.7 \pm 4.6$ & $298.19 \pm 1.83$ & $-351.76 \pm 0.71$ & $-105.56 \pm 1.87$ \\ 
G233-026 & $0.13 \pm 0.01$ & $13.63 \pm 0.03$ & $146.6 \pm 0.3$ & $230.93 \pm 0.48$ & $-245.18 \pm 0.6$ & $-15.79 \pm 0.04$\\ 
G262-021 & $0.23 \pm 0.01$ & $26.72 \pm 0.31$ & $302.9 \pm 3.6$ & $-328.81 \pm 1.46$ & $-237.36 \pm 0.48$ & $-119.68 \pm 0.31$\\ 
... & ...   & ...   & ... & ... & ... & ... \\

\hline 

\end{tabular}
\end{table*}}

{\scriptsize
\begin{table*}
\caption{A sample of additional kinematic parameters used to classify the stars as being dynamically coincident with the \textit{Gaia}-Sausage (Saug.) and \textit{Gaia}-Sequoia (Seq. G1 and Seq. G2) events. The first section shows  stars spectroscopically studied in this work. Errors were determined via 100 MC realizations exploring the errors associated with the input Gaia parameters. The full table is included with the online supplementary material.
\label{tab:actionparameters}}
\begin{tabular}{ccllllll}
\hline

Star & Subgroup & $J_{\phi}$ ($L_{z}$) & $J_{\text{r}}$ & $J_{z}$ & $e$ &$Z_{\text{max}}$ &$E$ \\
    &   & [kpc km/s]    & [kpc km/s]    & [kpc km/s]    &  & [kpc]    & [km$^{2}$s$^{-2}$] \\
\hline
G037-037 & ... & $-530.47 \pm 43.09$ & $3041.0 \pm 400.58$ & $2773.21 \pm 50.35$ & $0.78 \pm 0.02$ & $57.22 \pm 5.5$ & $-85411.3 \pm 4019.1$ \\ 
G158-100 & Seq. G2 & $-3038.04 \pm 113.08$ & $2471.18 \pm 409.74$ & $485.35 \pm 27.55$ & $0.73 \pm 0.03$ & $24.18 \pm 1.72$ & $-89901.4 \pm 4321.8$ \\ 
G184-007 & Seq. G2 & $-2684.13 \pm 13.44$ & $1074.79 \pm 50.16$ & $261.27 \pm 4.92$ & $0.6 \pm 0.01$ & $11.15 \pm 0.34$ & $-111490.3 \pm 979.4$ \\
G189-050 & ... & $-924.49 \pm 5.35$ & $1515.37 \pm 28.66$ & $140.06 \pm 6.15$ & $0.839 \pm 0.002$ & $7.6 \pm 0.3$ & $-126442.1 \pm 776.8$ \\ 
G233-026 & Saug. & $-69.42 \pm 4.89$ & $1172.94 \pm 3.84$ & $1.51 \pm 0.01$ & $0.981 \pm 0.001$ & $0.305 \pm 0.001$ & $-156530.1 \pm 112.4$ \\ 
G262-021 & Saug. & $-120.49 \pm 4.11$ & $2137.36 \pm 23.83$ & $210.23 \pm 1.19$ & $0.983 \pm 0.001$ & $11.16 \pm 0.17$ & $-122127.9 \pm 550.6$ \\ 
... & ...   & ...   & ... & ... & ... & ... & ...\\
\hline 
\end{tabular}
\end{table*}}

\section{Tables of Abundances and Equivalent Widths }

\begin{table*}
\caption{Elemental abundances relative to Fe I ({[}X/Fe I{]}) determined using the model atmospheres given in Table \ref{tab:stellparam} and EW measurements shown in Table \ref{tab:lines}. The number of lines used for each measurement is given in brackets beside the abundance value. Abundance errors were determined by adding the line-to-line dispersion $\sigma_{\mathrm{EW}}$ in quadrature with the uncertainties imposed by the stellar parameter errors
($\sigma_{Teff}$, $\sigma_{\text{log}g}$, $\sigma_{\text{[Fe/H]}}$, and $\sigma_{\xi}$). Note that NLTE corrections have been applied to the [Na/Fe] abundances and hfs corrections have been applied to the [Ba/Fe] abundances.}
\label{tab:abund}
\begin{tabular}{crrrrrr}
\hline
Abundance 	& G037-037	& G158-100	& G184-007	& G189-050	& G233-026	& G262-021 \\
Solar Value	& ...			& ...			& ...			& ...			& ...			& ... \\
\hline
{[}Li/Fe{]}	& $1.40\pm0.19$ (1)	& ...	& ...	& ...	& ...	& ... \\
 $3.26\pm0.05$ & ... & ... & ... & ... & ... & ...  \\
{[}Na/Fe{]}		& $-0.46\pm0.19$ (2)	& $-1.02\pm0.15$ (1)	& ...	& $-0.61\pm0.15$ (1)	& $-0.52\pm0.15$ (1)	& $-0.27\pm0.16$ (2) \\
 $6.24\pm0.03$ & ... & ... & ... & ... & ... & ...  \\
{[}Mg/Fe{]} 	& $0.06\pm0.19$ (1)		& $0.11\pm0.15$ (3)		& $-0.07\pm0.20$ (3)	& $-0.12\pm0.20$ (3)	& $0.38\pm0.25$ (2)	& $0.23\pm0.16$ (2) \\
$7.60\pm0.04$ & ... & ... & ... & ... & ... & ...  \\
{[}Si/Fe{]} 	& $0.42\pm0.19$ (1)		& ...	& ...	& $0.28\pm0.15$ (2)	& $0.09\pm0.15$ (3)	& $0.17\pm0.16$ (3) \\
$7.51\pm0.03$ & ... & ... & ... & ... & ... & ...  \\
{[}Ca/Fe{]}		& $0.20\pm0.06$ (14) 	& $0.18\pm0.04$ (17)	& $0.20\pm0.07$ (24)	& $0.17\pm0.08$ (27)	& $0.25\pm0.04$ (27)	& $0.34\pm0.07$ (23) \\
$6.34\pm0.04$ & ... & ... & ... & ... & ... & ...  \\
{[}Sc II/Fe{]}	& ... 	& ... 	& $-0.09\pm0.14$ (1)	& $-0.10\pm0.15$ (2)	& $-0.04\pm0.15$ (2)	& $0.18\pm0.16$ (2) \\
 $3.15\pm0.04$ & ... & ... & ... & ... & ... & ...  \\
{[}Ti I/Fe{]} 	& $0.30\pm0.19$ (3) 	& $0.29\pm0.06$ (19)	& $0.23\pm0.11$	(32)	& $0.20\pm0.13$ (29)	& $0.27\pm0.07$ (15)	& $0.38\pm0.11$ (41) \\
$4.95\pm0.05$ & ... & ... & ... & ... & ... & ...  \\
{[}Ti II/Fe{]} 		& $0.37\pm0.05$ (7) 	& $0.38\pm0.04$ (8)		& $0.34\pm0.04$ (11)	& $0.29\pm0.04$ (16)	& $0.20\pm0.05$ (12)	& $0.44\pm0.05$ (18) \\
... & ... & ... & ... & ... & ... & ...  \\
{[}V/Fe{]} 	& ... 	& ... 	& $0.30\pm0.14$ (2) 	& ...	& $0.02\pm0.15$ (1)		& $0.30\pm0.16$ (3) \\
$3.93\pm0.08$ & ... & ... & ... & ... & ... & ...  \\
{[}Cr I/Fe{]} 	& $-0.14\pm0.19$ (2)	& $0.03\pm0.06$ (14)	& $0.15\pm0.10$ (17)	& $0.08\pm0.13$ (16)	& $0.04\pm0.06$ (11)	& $0.14\pm0.11$ (15) \\
$5.64\pm0.04$ & ... & ... & ... & ... & ... & ...  \\
{[}Cr II/Fe{]}		& ...	& ... 	& ... 	& $0.18\pm0.15$ (3)		& $-0.28\pm0.15$ (2)	& $0.26\pm0.16$ (3) \\
... & ... & ... & ... & ... & ... & ...  \\
{[}Mn/Fe{]}	& ... 	& $-0.37\pm0.15$ (2)	& $-0.21\pm0.14$ (4)	& $-0.26\pm0.15$ (3)	& $-0.22\pm0.15$ (3)	& $-0.19\pm0.16$ (3) \\
$5.43\pm0.05$ & ... & ... & ... & ... & ... & ...  \\
{[}Fe I/H{]}		& $-2.01\pm0.08$ (51)	& $-2.24\pm0.05$ (113)	& $-1.67\pm0.07$ (150)	& $-1.41\pm0.10$ (164)	& $-1.34\pm0.06$ (156)	& $-1.38\pm0.07$ (155) \\
$7.50\pm0.04$ & ... & ... & ... & ... & ... & ...  \\
{[}Fe II/H{]}		& $-2.06\pm0.04$ (8)	& $-2.29\pm0.03$ (6)	& $-1.70\pm0.05$ (10)	& $-1.48\pm0.05$ (14)	& $-1.38\pm0.04$ (10)	& $-1.35\pm0.06$ (11) \\
... & ... & ... & ... & ... & ... & ...  \\
{[}Co/Fe{]}		& ... 	& ... 	& ... 	& ... 	& $0.41\pm0.15$ (1)	& $0.13\pm0.16$ (1) \\
$4.99\pm0.07$ & ... & ... & ... & ... & ... & ...  \\
{[}Ni/Fe{]}	& $-0.215\pm0.19$ (1) 	& $0.06\pm0.15$ (4)		& $-0.07\pm0.04$ (17)	& $-0.11\pm0.07$ (18)	& $-0.08\pm0.04$ (19)	& $0.04\pm0.04$ (23) \\
$6.22\pm0.04$ & ... & ... & ... & ... & ... & ...  \\
{[}Cu/Fe{]}		& ... 	& ... 	& ... 	& ... 	& $-0.44\pm0.15$ (1)	& $-0.15\pm0.16$ (1) \\
 $4.19\pm0.04$ & ... & ... & ... & ... & ... & ...  \\
{[}Zn/Fe{]}		& ... 	& ... 	& $0.09\pm0.14$ (2) 	& $-0.15\pm0.15$ (2)	& $-0.12\pm0.15$ (2)	& $0.09\pm0.16$ (2) \\
$4.56\pm0.05$ & ... & ... & ... & ... & ... & ...  \\
{[}Y II/Fe{]} 	& ... 	& ... 	& ... 	& $-0.11\pm0.15$ (1) 	& $-0.02\pm0.15$ (1)	& $0.14\pm0.16$ (1) \\
$2.21\pm0.05$ & ... & ... & ... & ... & ... & ...  \\
{[}Ba II/Fe{]}	& $0.25\pm0.19$ (2)		& $0.17\pm0.15$ (2)		& $0.13\pm0.14$ (4)		& $0.00\pm0.15$ (3)		& $0.08\pm0.15$ (3)		& $0.07\pm0.16$ (5) \\
 $2.18\pm0.09$ & ... & ... & ... & ... & ... & ...  \\
{[}$\alpha$/Fe{]}	& $0.21\pm0.10$			& $0.23\pm0.07$			& $0.20\pm0.14$			& $0.17\pm0.15$			& $0.26\pm0.05$			& $0.36\pm0.06$\\
\hline
\end{tabular}
\end{table*}

\begin{table*}
\caption{A sample of the atomic data and equivalent width measurements for the lines used in this study to determine chemical abundances. Specifics of this line list are discussed in Section \ref{sec:ewmeas}. The full table is included with the online supplementary material.}
\label{tab:lines}
\begin{tabular}{ccccccccccc}
\hline
Element & Wavelength    & $\chi$    & log gf  &   G184-007   & G189-050    & G158-100 & G262-021 & G233-026    & G037-037 \\
--  & [\AA]   & [eV] &  --  & [m{\AA}] & [m{\AA}]   & [m{\AA}] & [m{\AA}] & [m{\AA}] & [m{\AA}] \\
\hline
Fe I 	& 4388.407	& 3.60	& -0.682	& ...	& ...	& ...	& ...	& 46.6 	& ... \\
Fe I 	& 4430.614  & 2.22	& -1.659	& ...	& ...	& ...	& ...	& 78.8	& ... \\  
Fe I	& 4442.339	& 2.22	& -1.255	& 118.9	& 102.0	& 70.1	& 130.1	& 104.1	& ... \\
Fe I	& 4443.194	& 2.86	& -1.043	& 75.7 	& 64.3	& 32.3	& 81.4	& 64.1	& ... \\
Fe I	& 4447.717	& 2.22	& -1.342	& 105.3	& 91.1	& 61.7	& 111.7	& 92.3	& 48.4\\
... & ...   & ...   & ...   & ...   & ...   & ...  & ...   & ...   & ...\\
\hline
\end{tabular}
\end{table*}

% Don't change these lines
%\bsp	% typesetting comment
% \label{lastpage}
\end{document}